\documentclass[acmlarge,screen,nonacm]{acmart}

\usepackage{caption}
\usepackage{graphicx}
\usepackage{float} 
\usepackage{subcaption}

\usepackage{etoolbox}

\usepackage{algorithm}
\usepackage{amsmath}
\usepackage{algorithmic}

\usepackage{multirow}

\AtBeginDocument{%
  }

\author{Qing An}
\email{qa4@rice.edu}
\affiliation{%
  \institution{Rice University}
  \city{Houston}
  \state{Texas}
  \country{USA}
}

\author{Divyanshu Pandey}
\affiliation{%
  \institution{Rice University}
  \city{Houston}
  \state{Texas}
  \country{USA}
}
\email{dp76@rice.edu}

\author{Rahman Doost-Mohammady}
\affiliation{%
  \institution{Rice University}
  \city{Houston}
  \state{Texas}
  \country{USA}
}
\email{doost@rice.edu}

\author{Ashutosh Sabharwal}
\affiliation{%
  \institution{Rice University}
  \city{Houston}
  \state{Texas}
  \country{USA}}
\email{ashu@rice.edu}

\author{Srinivas Shakkottai}
\affiliation{%
  \institution{Texas A\&M University}
  \city{College Station}
  \state{Texas}
  \country{USA}}
\email{sshakkot@tamu.edu}

\begin{document}

\title{Multi-Provider Resource Scheduling in Massive MIMO Radio Access Networks}

\begin{abstract}
An important aspect of 5G networks is the development of Radio Access Network (RAN) slicing, a concept wherein the virtualized infrastructure of wireless networks is subdivided into slices (or enterprises), tailored to fulfill specific use-cases. A key focus in this context is the efficient radio resource allocation to meet various enterprises' service-level agreements (SLAs). In this work, we introduce a channel-aware and SLA-aware RAN slicing framework for massive multiple input multiple output (MIMO) networks where resource allocation extends to incorporate the spatial dimension available through beamforming. Essentially, the same time-frequency resource block (RB) can be shared across multiple users through multiple antennas. Notably, certain enterprises, particularly those operating critical infrastructure, necessitate dedicated RB allocation, denoted as private networks, to ensure security. Conversely, some enterprises would allow resource sharing with others in the public network to maintain network performance while minimizing capital expenditure. Building upon this understanding, the proposed scheduler comprises scheduling schemes under both scenarios: where different slices share the same set of RBs, and where they require exclusivity of allocated RBs. We validate the efficacy of our proposed schedulers through simulation by utilizing a channel data set collected from a real-world massive MIMO testbed. Our assessments demonstrate that resource sharing across slices using our approach can lead up to 60.9\% reduction in RB usage compared to other approaches. Moreover, our proposed schedulers exhibit significantly enhanced operational efficiency, with significantly faster running time compared to exhaustive greedy approaches while meeting the stringent 5G sub-millisecond-level latency requirement.

\end{abstract}


\settopmatter{printacmref=false}
\maketitle

\section{Introduction}
\label{sec:intro}

Massive MIMO has been an important part of the 5th Generation (5G) mobile network rollout, which has been in progress since the early 2020s. This technology significantly enhances overall network throughput by leveraging large-order multi-user MIMO (MU-MIMO) enabled through advanced beamforming techniques. However, the capital expenditure (CAPEX) associated with deploying Massive MIMO arrays is considerably higher than the lower-order MIMO base stations (BS) used in 4G and earlier 5G networks. This increased CAPEX necessitates meticulous planning by mobile network operators (MNOs) to ensure a balanced cost-benefit ratio for Massive MIMO deployment. Therefore, how to fulfill as many as users' service requirements with limited physical infrastructure is a crucial challenge for MNOs. Standing at the customers' points, based on their required types of services, such as throughput, latency or massive connectivity, they have diverse bandwidth demands. Meeting their network needs with the least amount of resources and costs is their top concern. Network slicing emerges as a solution to this challenge by enabling the partitioning and sharing of network resources across multiple slices or enterprises on a single physical infrastructure. Each enterprise typically has distinct use cases. For instance, entities with critical infrastructure or mission-critical applications, such as hospitals and university campuses, prioritize security and prefer not to share radio resources with others. These enterprises opt for private networks, which are accessible only to a specific set of devices. Furthermore, in order to save costs, most enterprises choose dependent private networks ~\cite{samsung}, wherein MNOs allocate spectrum to the enterprise based on cost and spectrum availability and are responsible for network deployment and maintenance. Conversely, enterprises in sectors such as retail and home networks, which do not have significant security concerns, are more willing to share radio resources with others to fulfill their service-level agreements (SLAs) cost-effectively. Such enterprises can be deployed in public networks, where they share the same resource blocks (RBs) with users from multiple enterprises. 

Previous work~\cite{flare,nvs} proposed the allocation of different RBs in LTE and 5G to various slices. Recent advancements, such as RadioSaber~\cite{radiosaber}, have introduced a channel-aware allocation of RBs to each slice to maximize network throughput. RadioSaber's architecture includes an inter-slice scheduler that polls different slices about their channel quality on each resource block and allocates them to slices that can achieve the highest rates. This approach separates the user allocation decision-making from the overall slice decision-making, allowing each intra-slice scheduler to manage its own users on the allocated RBs.
However, the challenges in designing a channel-aware RAN slicing with massive MIMO networks are much greater than Single-Input Single-Output (SISO) system considered in earlier works, such as RadioSaber. These challenges are two-fold: First, beamforming to multiple users must consider the channel correlation among the users. In the case of highly correlated channels, beamforming to multiple users will result in high inter-user interference, which in turn leads to poor throughput performance. Thus, any scheduler must take the correlation among users into account and interaction between the inter-slice scheduler and the intra-slice scheduler cannot simply be limited to the best channel quality in each slice. Second, the computational complexity of performing separate inter-slice and intra-slice schedulers increases exponentially with the number of RBs, slices, and UEs per slice. Such high-complexity solutions fail to achieve the sub-millisecond latency requirement for 5G and beyond networks. Thus, a scheduling framework is much needed wherein the intra- and inter-slice scheduling are performed jointly while still maintaining each individual slice's SLA guarantees and privacy requirements. In addition to being channel-aware, the scheduler must also be SLA-aware. This implies that the allocation of RBs should be guided by the specific SLAs of the enterprise to utilize the minimal number of RBs necessary to satisfy the enterprise's SLA requirements. By optimizing RB usage, additional resources can be freed to support more users or to facilitate other transmission needs, such as control channel signals and sensing signals. This approach yields reciprocal benefits for both enterprises and network operators.

\begin{figure}[htp]
  \centering
  \includegraphics[width=0.78\textwidth]{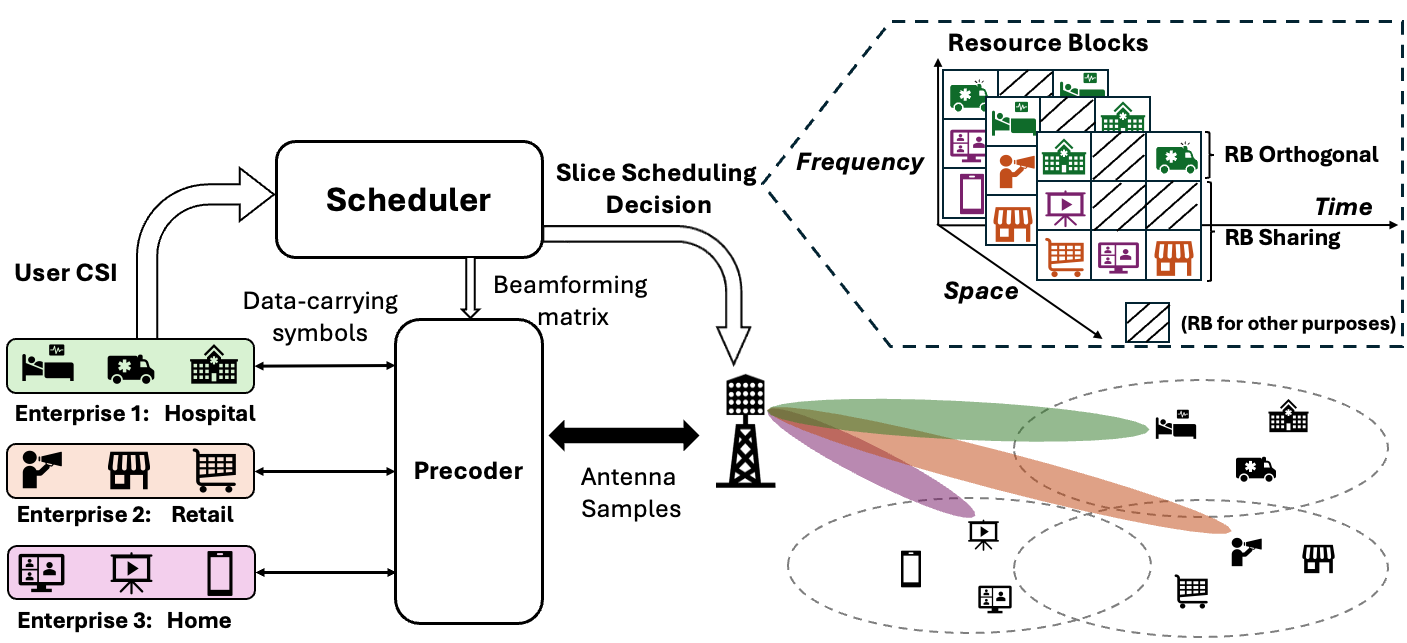}
  \caption{RAN Slicing based scheduling architecture in massive MIMO networks.} 
  \label{fig:intro}
\end{figure}

In this paper, we present a RAN slicing-based scheduling framework for massive MIMO networks which is both SLA-aware and channel-aware. The framework architecture is depicted in Fig.~\ref{fig:intro}.
The proposed scheduler not only allocates RBs to slices and users in a SLA-aware and channel-aware manner but also extends the allocation to spatial resources (beams), thereby going beyond the frequency-time resources considered in earlier works. It exploits the opportunity to allocate the same RB to multiple UEs simultaneously by leveraging multiple spatial dimensions available in massive MIMO. Two distinct use-case scenarios emerge in such cases. First, when each slice does not want an RB allocated to itself to be shared with UEs of other slices, such as in private networks where data security and privacy are crucial. Second, when slices only want assurance of their SLAs being met while being acceptable of resource sharing across slices, such as in public networks. Henceforth, we refer to the former as the RB-orthogonal case and the latter as the RB-sharing case. The orthogonality of RBs ensures that any given RB is allocated exclusively to a specific slice in any scheduling instant, and multiple UEs within that same slice can only use it. RB sharing ensures that UEs from multiple slices can simultaneously be allocated the same RB. In this work, we propose scheduling algorithms for both RB orthogonal and RB sharing. Since our primary objective is cost-saving for operators and enterprises, the scheduling problem posed in this work aims to minimize the number of resource blocks required to meet the SLA guarantees of each slice. The main contributions of the paper are as follows:
\begin{itemize}
    \item First framework for massive MIMO RAN slicing with SLA guarantees with the goal of optimizing resource usage, e.g. RBs.
    \item Both RB-orthogonal and RB-sharing algorithms are proposed with different levels of slice autonomy.
    \item Proposed algorithms perform close to optimal when compared to exhaustive search methods while offering significantly lower computational complexity making them compliant with stringent 5G latency requirements. 
    \item An exhaustive evaluation of the proposed schedulers on a real-world massive MIMO channel dataset under different network size configurations, correlation cases, mobility scenarios, and SLA constraints highlights the trade-offs between different schemes.
\end{itemize}

\section{Background and Related Work}
\label{sec:bg}
In this section, we provide a brief background on massive MIMO and its adoption in the 5G standard. We also discuss the problem of RAN slicing and its challenge as it pertains to massive MIMO. A literature review of RAN slicing and massive MIMO resource scheduling is also presented.

\textbf{Massive MIMO beamforming.} In theory, massive MIMO is referred to a cellular base station with an unlimited number of antennas. In~\cite{tenmyths}, it is shown that with $M$ antennas at the base station and $N$ single-antenna users, as the ratio $M/N$ goes to infinity, and when the channel is considered Gaussian, the beamforming matrix that maps the users’ data to antenna is simply the conjugate of the $M\times N$ channel matrix  $\textbf{H}$ to a scale. In practice, however, there is a limit to the number of antennas at a massive MIMO base station. The current deployment of massive MIMO in 5G is limited to 64 to 128 antennas~\cite{64_128}. The beamforming operation is typically done through Zero Forcing (ZF) or its regularized variant. The ZF beamformer is calculated as: $ \textbf{W} = \textbf{H}(\textbf{H}^H\textbf{H})^{-1}$. The channel matrix is estimated through uplink pilot signals transmitted by each UE to the base station. In the time-division duplex (TDD) regime, the same estimated channel matrix is often used for both uplink and downlink due to channel reciprocity. 

\textbf{Numerology of 5G Frame Structure.} 5G cellular networks operate based on OFDM signaling, where data is transmitted across frequency subcarriers. The 5G frame format includes 10 ms frames with slots of typically 1 ms length and 14 OFDM symbols~\cite{frame}. The number of subcarriers in each OFDM symbol is a function of the system bandwidth. The subcarriers across all OFDM symbols of a slot are grouped into resource blocks (RB), each containing 12 subcarriers. For e.g., in a 100 MHz 5G system, there are 273 resource blocks. A special pilot signal known as the sounding reference signal (SRS) is used to obtain the channel estimate for beamforming in massive MIMO~\cite{srs}. Through the channel estimate obtained from the SRS pilots, the massive MIMO 5G base station can schedule multiple users in each resource block for data transmission or reception through multi-user beamforming. In other words, resources are scheduled across all time, frequency, and spatial domains. 

\textbf{Massive MIMO Resource Scheduling.} The resource scheduling problem in massive MIMO is significantly more complex than in SISO systems. In the SISO case, the scheduler can allocate an RB to a user with the highest rate. 
In massive MIMO, however, the rate-optimal allocation of multiple users to each RB depends on how much rate the groups of users can collectively achieve. Therefore, a rate-optimal scheduler should solve a combinatorial problem that considers all combinations of users. The same goes with the proportional-fair or max-min fair algorithms that try to provide fairness along with maximizing rate. Such combinatorial problems are provably NP-hard and clearly not feasible to run in massive MIMO networks of 5G, which require very low latency in scheduling decisions~\cite{mcore}. 
Many related work propose sub-optimal heuristic-based MU-MIMO schedulers~\cite{chen2020twc,liu2017systems,900644,4621387,6803179}. While they try to strike a balance between complexity and performance, their complexity does not scale to large networks or they significantly underperform the optimal scheduling policies. With the aid of prompt inference of AI models, several ML-models are proposed~\cite{guo2020globecom,chen2021eusipco,shi2018wcsp,bu2019iccc,chen2021joint,lopes2022deep,10038693}, which leverage deep reinforcement learning for optimal user selection, yet their applicability across diverse traffic scenarios remains limited. 
The aforementioned traditional schedulers typically prioritize the distinct requirements of individual users over accommodating groups of users with similar Quality of Service (QoS) needs. In the context of 5G, there is a growing emphasis from operators and enterprises on addressing enterprise-level QoS requirements \cite{ns}. In contrast to conventional approaches that adopt a uniform resource allocation strategy, the proposed scheduler facilitates the customization of network behavior to accommodate various service types, including RB orthogonality or sharing. This adaptability is essential for meeting the diverse performance demands of modern 5G services.

\textbf{Massive MIMO RAN Slicing.}
RAN slicing has emerged as a prominent strategy for virtualizing radio resources, enhancing cost and power efficiency by accommodating multiple service slices within the network \cite{ran_slice}. Network Virtualization Substrate (NVS) \cite{nvs} allocates all RBs within a TTI to a single slice, employing a weighted round-robin approach to meet each slice's target throughput requirements specified in the SLA. However, NVS lacks channel awareness, leading to suboptimal resource allocation decisions. Advanced RAN virtualization frameworks like Orion and Scope \cite{orion,scope} build upon NVS, aiming to enhance resource allocation efficiency. RadioSaber \cite{radiosaber} further refines this approach by incorporating channel awareness into the slice-level scheduler, optimizing RB allocation based on users' Channel Quality Indicators (CQIs). Ensuring slice-level service quality is a primary concern across various RAN slicing proposals \cite{9490293,8385093,9076124}. Recent efforts, such as Zipper \cite{zipper}, a ML-based algorithm to compute SLA-compliant schedules in real-time, albeit with a focus on application-level service assurance. Despite these advancements, existing RAN slicing frameworks predominantly focus on single-antenna systems and commonly employ slot-based and RB-based slicing methodologies \cite{nvs, orion, radiosaber}.
In summary, the slicing management framework guarantees SLA compliance by allocating resources such as slots or RBs to different slices based on metrics like average throughput or latency. Similar concepts can be used in massive MIMO RAN slicing. However, spatial streams represent a third dimension, alongside time and frequency dimensions. While allocating spatial streams in each RB to different slices enhances cost/power efficiency by resource reuse, spatial stream slicing is less effective than slot and RB slicing due to residual inter-user interference based on the effectiveness of the beamforming procedure. This interference can impact operator services. Additionally, slices utilizing spatial streams in shared RBs must employ a common precoder, conceding some functionality to the slice management framework.
In \S\ref{sec:problem}, we discuss massive MIMO RAN slicing and the design space in more detail.
\section{Problem Overview}
\label{sec:problem}

The design space of the proposed scheduler includes a scheduling scheme accommodating various slices, each with distinct SLA requirements and users. It supports the operation of both private and public 5G network functionalities on a single base station, employing either cooperative or exclusive resource allocation across slices. This framework considers two scenarios for massive MIMO RAN slicing: RB-orthogonal, where a single RB is dedicated to a single slice, and RB-sharing, where a single RB is shared among multiple slices. Both scenarios aim to satisfy the SLAs of all slices while minimizing RB usage.


\subsection{RB-Orthogonal: Problem Formulation} In this case, any given RB can only be allocated to users from the same slice. This constraints the optimal usage of each RB; however, it might be necessary due to security and privacy issues that stem from resource sharing across different slices.

\textbf{Inter- and Intra-slice scheduler :} Within the RB-Orthogonal case, we can have two situations: First, separate inter- and intra-slice schedulers make decisions at slice and user levels, respectively. To do this, the intra-slice scheduler needs to let the inter-slice scheduler know which user it intends to serve if given a certain RB, depending on its own private policy. The inter-slice scheduler allocates RBs to various slices, while the intra-slice scheduler assigns these RBs to individual users within each slice. Second, where there is a single scheduler that directly takes decisions at the user level while maintaining the slice SLAs. Note that having a separation between inter- and intra-slice schedulers allows a level of private autonomy for each slice (enterprise) to decide its own scheduling policy. The intra-slice scheduler can select users based on any customizable policy such as Proportional Fairness (PF) or Maximum Rate (MR). The inter-slice scheduler only schedules a slice depending on the response to its query from the slice and does not dictate the scheduling at the user level. However, an exhaustive search at the intra-slice level is often required to respond to the inter-slice scheduler's query. Thus, computationally, it is more efficient to let a single scheduler directly take a decision at the user level.The latter approach can also accommodate various scheduling policies concurrently.

Let $\mathcal{B},\mathcal{S}$ denote the set of RBs and slices respectively. Every slice $s \in \mathcal{S}$ contains a distinct set of users denoted as $\mathcal{K}_s$ such that $\sum_s |\mathcal{K}_s| = N$, where $N$ is total number of users. Let us denote $x_{k}^{b,t}$ and $x_{s}^{b,t}$ as the user-level and slice-level binary decision variables for user $k$ on RB $b$ respectively, such that
$
    x_{s}^{b,t} = 1,  \exists\ k\in \mathcal{K}_{s}, \text{s.t.}\ x_{k}^{b,t} = 1
$.
Subsequently, the scheduling problem for the RB-orthogonal case can be written as the following optimization problem:
\begin{equation}\label{eq:RB_Orth_Opt}
\begin{alignedat}{2}
  \min_{x \in \{0,1\}}  \, \mathop{\mathbb{E}_t}\left[ \sum_{s\in \mathcal{S}}\sum_{b\in \mathcal{B}}x_{s}^{b,t} \right] \quad \quad
  \text{s.t.} \,\, & \sum_{s \in \mathcal{S}}x_{s}^{b,t} \le 1 , \forall b, \forall t \quad \text{and} \quad \mathop{\mathbb{E}_t}\left[\sum_{b\in \mathcal{B}}r_{s}^{b,t}\right] \ge \gamma_{s} , \forall s
\end{alignedat}
\end{equation}
where the optimization variable $x$ denotes $x_s^{b,t}$ if we have a separate inter- and intra-slice scheduler, and $x$ denotes $x_k^{b,t}$ when there is a single scheduler taking decisions at the user level. The objective function remains the same in both cases. The entity $r_{s}^{b,t}$ is the achieved data rate of slice $s$ on RB $b$ at TTI $t$ and $\gamma_{s}$ is the SLA in terms of minimum throughput guarantee for slice $s$ on average across several TTIs. The notation $\mathbb{E}_t[\cdot]$ denotes average across $t$. Further, at any given TTI instant $t'$, SLA deficit of slice $s$ can be computed as:
\begin{equation}
\label{eq:delta}
    \Delta^{t'}_{s} = \max\big\{0, t'\times \gamma_{s} - \sum_{t=0}^{t'-1}\sum_{b \in \mathcal{B}}r_{s}^{b,t}\big\}
\end{equation}
The objective function in Eq.~\eqref{eq:RB_Orth_Opt} is to minimize the average allocated resource blocks to all the slices across time. The first constraint in Eq.~\eqref{eq:RB_Orth_Opt} imposes the orthogonality of RB allocation, and the second constraint ensures that the SLA requirement is met for all slices.
Such an optimization problem can be abstracted as binary integer programming (BIP), which is NP-complete. Some heuristic methods are proposed to solve BIP, such as the Branch and Bound (B\&B) method~\cite{branch}. However, it is impeded by large computational complexity, especially as the number of binary variables and constraints increase~\cite{bb_complexity}. 

\subsection{RB-Sharing: Problem Formulation} RB-sharing maximizes the utilization of RBs by sharing it with users from multiple slices. Thus, there is no notion of a separation between inter- and intra-slice schedulers in this case.
The corresponding optimization problem for the RB-sharing case can be formulated as:
\begin{equation}
\begin{alignedat}{2}
  \min_{x_{k}^{b,t} \in \{0,1\}}  \quad & \mathop{\mathbb{E}_t}\left[\sum_{b \in \mathcal{B}}\min(1,\sum_{k=1}^{N}x_{k}^{b,t}) \right] \quad \quad
  \text{s.t.} \quad & \mathop{\mathbb{E}_t}\left[\sum_{b \in \mathcal{B}}r_{s}^{b,t}\right] \ge \gamma_{s}, \forall s.
\end{alignedat}
\end{equation}
This objective function is also a combinatorial problem and NP-hard. It is more computationally complicated than RB-orthogonal because user selection on each RB extends from users within the slice to all users in the network, which makes the search space exponentially larger. 


\section{Scheduler Design}
\label{sec:method}

\begin{figure}[htp]
  \centering
  \includegraphics[width=0.78\textwidth]{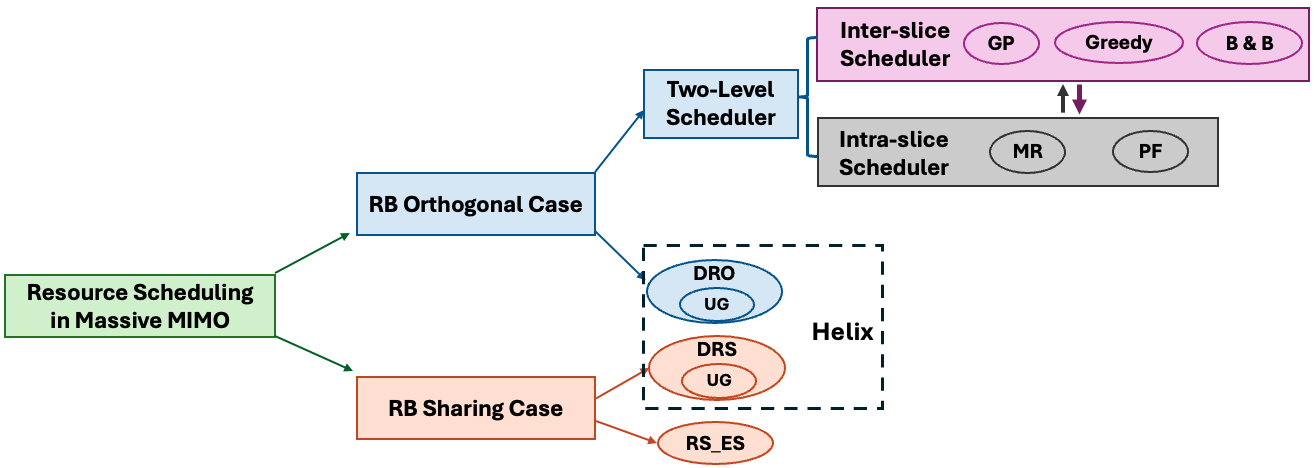}
  \caption{Overview of implemented scheduling algorithms.} 
  \label{fig:alg}
\end{figure}

To minimize the cost for both operators and enterprises, we align the objectives of scheduler design in both RB-orthogonal and RB-sharing scenarios to focus on fulfilling SLAs of slices while minimizing the consumption of resource blocks. Consequently, the scheduler must be both channel-aware and SLA-aware, as discussed in~\S\ref{sec:intro}. Additionally, maintaining low computational complexity is another design criterion, as emphasized in~\S\ref{sec:problem}, such that the scheduler can be implemented within each TTI. Consequently, we propose a novel scheduling scheme that includes scheduling algorithms for both RB-orthogonal and RB-sharing. An overview of implemented scheduler algorithms in this work is depicted in Fig.~\ref{fig:alg}. Notably, resource allocation in SISO networks relies on RB-orthogonality, as each RB is only occupied by a single user. Therefore, in~\S\ref{sec:exps}, we adapt~\cite{radiosaber} for MIMO networks, utilizing it as a baseline in the RB-orthogonal scenario.

\subsection{User Grouping (UG)} \label{sec:user_grouping} 
Since our proposed scheduler aims to schedule multiple users on the same RB simultaneously by using different spatial streams, an appropriate grouping of users based on inter-user correlation is required to guide the scheduler design. Thus, as a prelude to our proposed scheduling algorithms, we first present a user-grouping approach. Suppose $\mathcal{N}$ denotes the set of all users. In that case, the objective of user-grouping is to partition $\mathcal{N}$ into disjoint sets $\{ \mathcal{N}_1, \mathcal{N}_2,\dots,\mathcal{N}_{\eta} \}$ such that $\eta$ is minimum and users within any $\mathcal{N}_i$ have low correlation. The inter-user correlation between user $i$ and $j$ can be calculated using the CSI vectors as stated in \cite{yang2018spawc}. 
Accordingly, a size $N \times N$ binary user correlation matrix $\mathbf{G}$ can be generated by defining a correlation threshold $c_{th}$ ~\cite{yang2018spawc}. If $c_{i,j} > c_{th}$, the $(i,j)$th element of the correlation matrix, $\mathbf{G}_{i,j} = 1$  otherwise, $\mathbf{G}_{i,j} = 0$. The diagonal values of this matrix are all set to 0. This binary user correlation matrix can then be represented as a graph where each vertex represents a user and a 1 in the $(i,j)th$ index of the correlation matrix corresponds to an edge between the $i$th and $j$th vertices of the graph. Since we want to form groups of users with low correlation, the graph representation allows us to solve user grouping as a graph coloring problem~\cite{graph}. Graph coloring involves assigning colors to the vertices of a graph using a minimum number of colors such that no two adjacent vertices share the same color. The vertices (users) with the same colors can be grouped together, the number of colors denote the number of user groups $\eta$. Graph coloring is a well-researched topic for which several algorithms are proposed in the literature ~\cite{rokos2015fast, graphcolor}. With the help of parallel computing techniques, these algorithms are efficient and scalable to handle thousands of vertices. Since user grouping is a function of correlation statistics, it may not significantly vary across different $b$ and $t$ specially with slow time-varying channels. Hence, depending on practical considerations, computing such user grouping only once, or updating it only after several TTIs should suffice.


\subsection{Scheduler for RB-Orthogonal case}
In this section, we propose two algorithms: Greedy Plus (GP) and Delta Algorithm for the RB-Orthogonal (DRO). Greedy Plus (GP) is an inter-slice scheduler that operates separately from the intra-slice scheduler. DRO directly takes scheduling decisions at the user level without separating inter- and intra-slice schedulers, thereby avoiding an exhaustive search by any intermediary. Note that since Eq. \eqref{eq:RB_Orth_Opt} can be formulated as a binary integer programming (BIP) problem, we employ Gurobi optimizer~\cite{gurobi} to numerically provide an optimal solution using the Branch and Bound (B\&B)~\cite{branch}. The complexity of solvers such as Gurobi is too high to meet the stringent latency requirements of 5G networks. Consequently, we will use the optimal solution generated by Gurobi only to evaluate our algorithm in~\S\ref{sec:exps} as a benchmark method.

\subsubsection{Greedy and Greedy Plus (GP) Approaches }
To make the scheduling process channel-aware, the intra-slice scheduler responds to the following query by the inter-slice scheduler: which user will receive a given RB if it is allocated to the slice by the inter-slice scheduler?  Based on the response from such a query, the inter-slice scheduler in~\cite{radiosaber} greedily allocates RBs while prioritizing slices with favorable channel quality and ceases allocation once the RB quota is met. Such a greedy inter-slice scheduler, as proposed in ~\cite{radiosaber}, makes locally optimal decisions at each step, which may fail to satisfy the long-term SLAs. 

\begin{figure}[htp]
  \centering
  \includegraphics[width=0.8\textwidth]{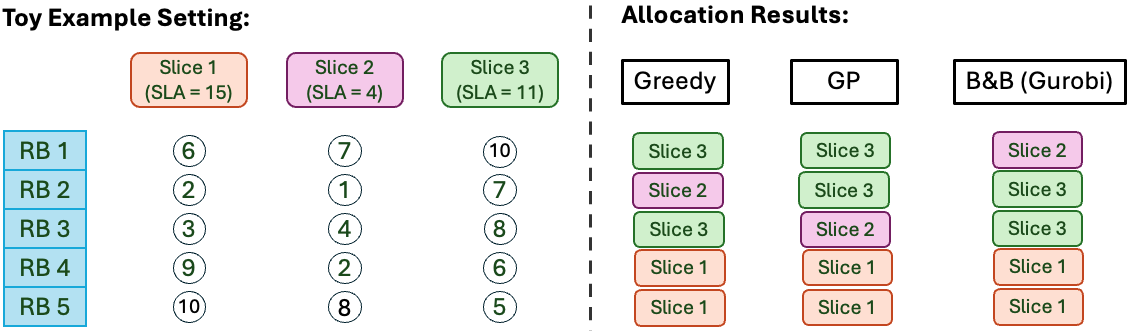}
  \caption{Illustrative Example Comparing Greedy, Greedy Plus, and Gurobi: Consider 3 slices and 5 RBs. The achieved data rates by intra-slice schedulers are shown in circles. \textit{Greedy} - $RB5 \rightarrow S1$,  $RB1\rightarrow S3$, $RB4\rightarrow S1$, after which $S1$ has exceeded its SLA. Then, $RB3\rightarrow S3$, $RB2\rightarrow S2$. \textit{GP} - Sort RBs in descending order within each slice, then serve the slice with largest SLA, thus $RB5 \rightarrow S1$, remove RB5 from lists of other slices and update SLA deficit of S1. Next S3 has largest deficit, thus $RB1 \rightarrow S3$. Further, $RB4\rightarrow S1$, completing S1's SLA requirement. Further, $RB5 \rightarrow S1$, $RB3\rightarrow S2$, and $RB2 \rightarrow S3$. \textit{Gurobi}-employs Branch and Bound method.} 
  \label{fig:toy}
\end{figure}

An improvement of the Greedy approach, which we refer as Greedy Plus (GP), recognizes average SLA deficits in Eq.~\eqref{eq:delta} and allocate the most suitable RBs to each slice to meet SLAs. Thus GP is cognizant of scenarios where assigning the RB that greedily provides the highest data rate to a user is not recommended if the corresponding SLA deficit is too small. GP sorts RBs based on achievable data rates within each slice rather than globally and starts allocation with the slice with the largest SLA deficits. After each RB allocation, it updates the SLA deficits using Eq.~\eqref{eq:delta} and removes the allocated RB to a slice from the sorting lists of other slices. The scheduling of a slice will conclude once its SLA is met. An illustrative example is presented in Fig \ref{fig:toy} showing that Greedy approach can fail to meet the SLA of each slice while over-serve a few slices with good channel. GP ensures no SLA violations, and performs similar to Gurobi.  

\subsubsection{{Delta RB Orthogonal (DRO)}} Performace of any scheduling scheme with separate inter- and intra-slice schedulers depends on how quickly the intra-slice scheduler can respond to inter-slice scheduler's query regarding the best user. In SISO networks, this is feasible within a short time since the best user can be found using CQI~\cite{radiosaber}. However, in MIMO networks, CQI is not as reliable because CQI acquisitions for beamformed users are independent, meaning each user may not be aware of the number of other MU-MIMO layers being used. Consequently, CQI feedback lacks context, which is even more problematic in massive MIMO networks~\cite{cqi}. To accurately estimate the achieved data rate of each potential user set on an individual RB, Shannon capacity formula~\cite{capacity_cal} can be used. We propose a low-complexity method that identifies the user combination with the highest potential to achieve the maximum rate on each RB without performing an exhaustive search.

Our proposed approach selects a user set by evaluating user channel gain and inter-user correlation, ensuring that RBs are allocated to users with low correlation and good channel quality. To this end, user-grouping proposed in \S\ref{sec:user_grouping} can be used. Additionally, we prioritize scheduling for slices with larger SLA deficits. Let us refer to the set of all slices with non-zero SLA deficit as active slice, denote it as $\mathcal{S}_{act}^t$, and the union of the set of all users in $\mathcal{S}_{act}^t$ as $\mathcal{K}_{act}$. Then we partition this set into two sub-sets of users $\mathcal{K}_{act}^t = G_l^t \cup G_s^t$ defined as:
\begin{equation}
\label{eq:class}
G_l^t= \{ i: \Delta_{slice(i)}^t \geq \Delta_{avg}^t \} \text{ and } G_s^t = \{ j:  \Delta_{slice(j)}^t < \Delta_{avg}^t \}, \text{ where } \Delta_{avg}^t = \frac{\sum_{s \in \mathcal{S}_{act}}\Delta_s^t}{|\mathcal{S}^t_{act}|}.   
\end{equation}
The function $slice(i)$ denotes the slice of user $i$, and the quantity $\Delta_{avg}^t$, denotes the average of all non-zero $\Delta_s^t$ from Eq.~\eqref{eq:delta}. Further, $G_l^t$ and $G_s^t$ denote subset of slices with more than and less than average SLA deficit respectively, making the users within $G_l^t$ the higher priority set of users. From $G_l^t$, we pick the user-RB ($k',b'$) pair with the highest channel gain as:
\begin{equation}
\label{eq:max_cg}
(k',b') =  \underset{k \in G_l^t,b \in \mathcal{B}}{\text{arg max}}\quad ||\mathbf{h}^{b,t}_{k}||^2.
\end{equation}
This allows us to initiate allocation from RB $b'$ and user $k' \in G_l^t$ where $slice(k')=s'$.
Given that this resource allocation occurs in massive MIMO networks, and we want to ensure RB orthogonality, we must also select other users from the same slice $s'$. As discussed in~\cite{tenmyths}, increasing the number of scheduled users $K$ in massive MIMO is not always beneficial, and there is an optimal $M/K$ ratio for maximizing spectral efficiency. 
While our proposed algorithm applies to any generic $K<N$ value, for implementation, $K$ can be determined following the guidelines in~\cite{tenmyths}. Thus, after finding a user $k'$ in slice $s'$ to be served, we need to find $K-1$ additional users, which should have a low inter-user correlation from slice $s'$ to schedule with user $k'$ on RB $b'$. Therefore, we group users into clusters where users exhibit low correlation with each other using the technique proposed in~\S\ref{sec:user_grouping}. We then identify the group containing user $u'$ and select the top $K-1$ users with the highest channel gain from the same slice $s'$ for scheduling. If the group contains less than $K$ users, we find user $j$ with the highest channel gain on RB $b'$ from other groups and continue selecting uncorrelated users from user $j$'s group until we have total $K$ users or all users in slice $s'$ are selected.
This approach may introduce some inter-user interference, but an increased number of scheduled users will compensate for it on data rate as long as the total number is smaller than $K$. After identifying the user set for scheduling on RB $b'$, we can use Shannon capacity equation~\cite{capacity_cal} to estimate the total achieved rate of slice $s'$ and update $\Delta_{s'}^t$. The scheduler will continue to use the updated SLA deficits to select subsequent RBs and corresponding user sets until all slice SLAs are met. The detailed algorithmic implementation is provided in Algorithm \ref{alg:Helix}.

\begin{algorithm}[htb]
\caption{Algorithmic Description of DRO and DRS approaches}
\label{alg:Helix}
\begin{algorithmic}[1] 
\REQUIRE ~~ 
    RB set $\mathcal{B}$; Slice set $\mathcal{S}$; Channel $\textbf{H}^{b,t}$; Users $K$; TTIs $T$; Decision to run DRO or DRS.
\ENSURE ~~ 
    Average number of allocated RBs
    \STATE $RB\_alloc\_total \gets 0 ${\COMMENT{Total number of allocated RBs}};
    \FOR{$t=0$; $t<T$; $t++$}
        \STATE $\left\{\mathcal{N}_1, \mathcal{N}_2, ... \mathcal{N}_{\eta}\right\}^{b,t} \longleftarrow user\_grouping(\mathbf{H}^{b,t})$ {\COMMENT{user grouping as per \S\ref{sec:user_grouping}}} 
        \STATE Update $\Delta_{s}^{t}, \forall s$ using Eq.~\eqref{eq:delta} and $\mathcal{B}^t\leftarrow \mathcal{B}$ {\COMMENT{update SLA deficits}}  
        \WHILE {$(\exists \Delta_{s}^{t} > 0 )$\ and\ $(len(\mathcal{B}^t) > 0)$}
            \STATE $\text{sel\_UE} \gets [ ] ${\COMMENT{Empty selected UE list}}
            \STATE Find $G_{l}^{t}, G_{s}^{t}$, and $(k',b')$ using Eq.~\eqref{eq:class} and ~\eqref{eq:max_cg} {\COMMENT{large, small $\Delta$ groups, best User-RB pair}}
            \STATE Identify $\mathcal{N}_i^{b',t}$ such that $k'\in \mathcal{N}_i^{b',t}$ {\COMMENT{set of low-correlated users with user $k'$}}
            \IF{DRO}
                \STATE  $G_l^{t} \leftarrow \{k: slice(k)=slice(k')\}$ {\COMMENT{change $G_l^t$ to set of all users in slice with user $k'$}}
            \ENDIF
            \WHILE{$len(\text{sel\_UE}) < K$}            
                \STATE Set $\mathcal{K}_{large}\leftarrow \{k : k \in (\mathcal{N}_i^{b',t} \cap  G_{l}^{t})\}$ {\COMMENT{set of low-correlated users within $G_l^t$}}
                \STATE Sort $\mathcal{K}_{large}$ in descending order of channel gain
                \STATE $\text{sel\_UE}.append\left(\mathcal{K}_{l}\left[0:\min (K - len(\text{sel\_UE}), len(\mathcal{K}_{large}))-1\right]\right)$
                    \IF{($len(\text{sel\_UE}) < K$ \& DRS)}
                    \STATE $\mathcal{K}_{small} \leftarrow \{k : k \in (\mathcal{N}_i^{b',t} \cap G_s^t\} $ {\COMMENT{Pick from small delta group if DRS}}
                    \STATE Sort $\mathcal{K}_{small}$ in descending order of channel gain
                       \STATE $\text{sel\_UE}.append\left(\mathcal{K}_{s}\left[0:\min (K - len(\text{sel\_UE}), len(\mathcal{K}_{small}))-1\right]\right)$
                    \ENDIF
                \IF{$len(\text{sel\_UE}) < K$}
                    \STATE Find $k' \gets \underset{k \in G_l^t, k \notin \mathcal{N}_i^{b',t}}{\text{arg max }} ||\mathbf{h}^{b',t}_k||$, and set $\mathcal{N}_i^{b',t} \leftarrow \mathcal{N}_j^{b',t}$ s.t. $k' \in \mathcal{N}_j^{b',t}$  
                \ENDIF
            \ENDWHILE
            \STATE Update $\Delta^{t}_{s}$ using Eq.~\eqref{eq:delta}
            \STATE $RB\_alloc\_total ++$
            \STATE $\mathcal{B}^t \gets \mathcal{B}^t- \{b'\}$ {\COMMENT{remove selected RB $b'$ from available RB set at TTI $t$}}
        \ENDWHILE
    \ENDFOR 
\RETURN $RB\_alloc\_total/T$ 
\end{algorithmic}
\end{algorithm}

\subsection{RB Sharing Case}
RB-Orthogonal does not provide an efficient solution when the users of a single slice are high-correlated. For example, in cluttered environments where users from the same slice are concentrated in a single location, such as in a crowded shopping mall, high inter-user interference occurs due to their strong correlation, leading to reduced system performance. Besides, in scenarios with sparse slices that contain very few active users (i.e., the number of users is fewer than $K$), such as resident networks in late night hours, exclusively assigning RB to these slices results in inefficient resource utilization. The RB sharing approach addresses these issues by enabling radio resources to be shared among users from all slices, thus enhancing RB utilization. This method operates in a fully centralized manner and collaboratively determines the optimal user combination across slices. While RB sharing can achieve superior resource saving compared to the RB-orthogonal approach, it incurs a higher computational overhead due to the increased complexity of user and RB combinations as outlined in~\S\ref{sec:problem}. Consequently, performing an exhaustive search for the optimal solution in a large-size network is impractical, so we propose a scheduler with low computational complexity.

\subsubsection{Delta RB Sharing (DRS)}
Building on the principles of DRO, we introduce a low-complexity approach tailored for the RB sharing scenario. The core concept involves prioritizing slices with substantial SLA deficits and allocating RBs to users with high channel quality and low correlation. Initially, we classify slices based on the average SLA deficit $\Delta_{avg}$ of active slices, similar to the RB-orthogonal scenario. Next, we identify the user-RB pair with the highest channel gain to initiate the allocation process. We group users as explained in ~\S\ref{sec:user_grouping} and identify the user group containing the selected user and pick additional $K-1$ users in descending order of channel gain. Unlike the RB-orthogonal scenario, where users must be selected from the same slice, RB-sharing allows users to be selected from any slice. However, users from slices with larger deficits are given higher priority, and users from slices with smaller deficits are only selected if insufficient users are available in the larger deficit group. SLA deficit update follow the same procedure as in the RB-orthogonal scenario. The RB allocation is performed sequentially and continues until all SLAs are met. Implementation details are provided in Algorithm~\ref{alg:Helix}.

\subsubsection{Exhaustive search in RB-Sharing (RS\_ES)}
To compare the performance of DRS in RB sharing with a benchmark method, we implement a near-optimal method known as RS\_ES, which performs a user-level exhaustive search. Specifically, RS\_ES first identifies the initial $(k',b')$ user-RB pair, similar to DRS. However, for selecting the remaining $K-1$ scheduled users on the same RB, unlike selecting for a fixed user group, RS\_ES conducts an exhaustive search among all users to find the optimal set of UEs. It then updates the SLA deficit following the same procedure as in DRS, picks the next best RB-user pair, and continues the process till all SLAs are met. Note that performing an exhaustive search without picking an initial user-RB pair would be a computationally infeasible approach even for small-scale networks since the search space increases exponentially with both the number of resource blocks and the number of users. Thus, RS\_ES provides a near-optimal solution by partly introducing user-level exhaustive search within the DRS approach.

\subsection{RB Parallelism}\label{sec:RB_parall}
In both DRO and DRS, RB allocation occurs sequentially because the outcomes of previous RB allocations adjust the SLA deficits, thereby affecting the priority of slices in subsequent allocations. This process determines whether a slice is categorized into a large or small delta group for the next allocation round. However, this sequential allocation mechanism leads to increased execution time, potentially violating the stringent latency requirements of 5G networks, particularly in RB-demanding cases (i.e. tight SLA constraints). However, because we use average SLA deficits to do slice classification, a single RB allocation result most often doesn't significantly impact the average deficits, and thus, the slice classification remains unchanged. This allows us to allocate multiple RBs using the same slice classification. This observation enables us to perform multiple RB allocations in parallel, thereby reducing execution time. 

More specifically, in Algo \ref{alg:Helix}, by finding multiple user-RB pairs $(k',b')$ in step 7, steps 8 to 24 can be parallelly executed for all such pairs. 
The number of user-RB pairs we pick would determine the number of parallel threads required. One possible way to pick this degree of parallelism can be determined by the ratio of total SLA deficits of the current TTI to the average achieved data rate per RB in the previous TTI. The degree of parallelism can be dynamically adapted as per changing SLA-deficits and execution time requirements. We will evaluate the effectiveness of RB parallelism in~\S\ref{sec:exps}, demonstrating its impact on reducing running time with only a minor performance degradation.

\section{Experiments}
\label{sec:exps}
In this section, we comprehensively evaluate our proposed scheduler in massive MIMO-based trace-driven simulation under various network configurations. The highlights of our evaluation are as follows:
\begin{itemize}
  \item In RB-orthogonal case, GP's performance is near-optimal and better than the Greedy algorithm (\S\ref{sec:exp_per}) in small and medium-size networks. 
  \item DRO achieves comparable performance to GP and is scalable to real-world networks. (\S\ref{sec:exp_per}). 
  \item DRS provides near-optimal solution and consumes fewer RBs than any orthogonality-based schedulers due to the cooperative sharing of resources (\S\ref{sec:exp_per}).
  \item The benefits of our proposed scheduler hold in different network configurations (small, medium, and real-world size) and scenarios (static and mobility) (\S\ref{sec:exp_per}).
  \item Our proposed schedulers can support diverse scheduling policies and SLAs (\S\ref{sec:policy}).
  \item The running time of DRO and DRS can meet 5G sub-millisecond latency requirement with the help of RB parallelism (\S\ref{sec:exp_time}).
\end{itemize}

\subsection{Experimental Setup}
\label{sec:exp_sp}
\begin{figure}[t]
	\centering
	\begin{subfigure}{0.47\linewidth}
		\centering
		\includegraphics[width=0.95\linewidth]{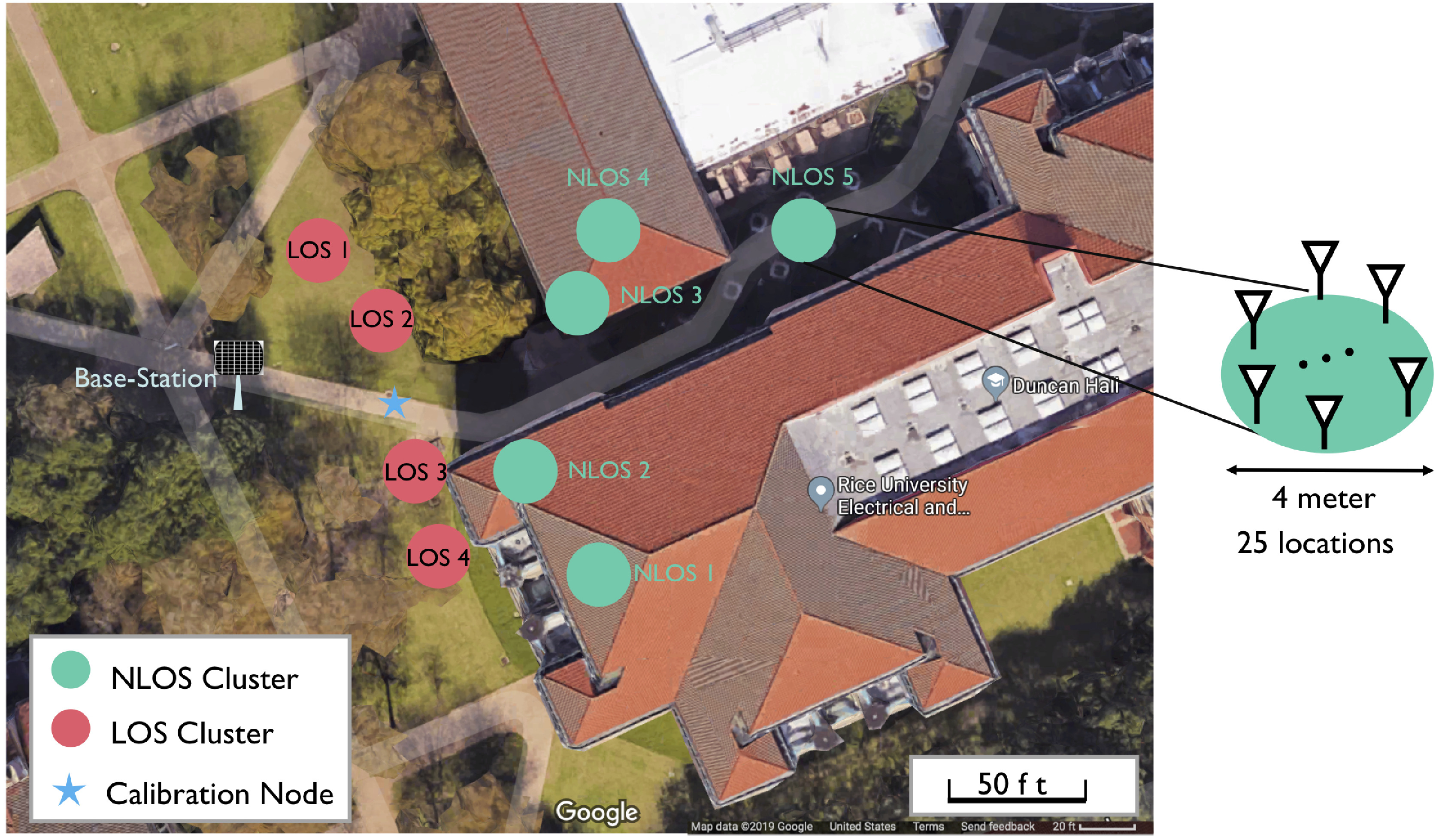}
            \caption{}
		\label{subfig:topo}
	\end{subfigure}
	\centering
	\begin{subfigure}{0.33\linewidth}
		\centering
		\includegraphics[width=0.95\linewidth]{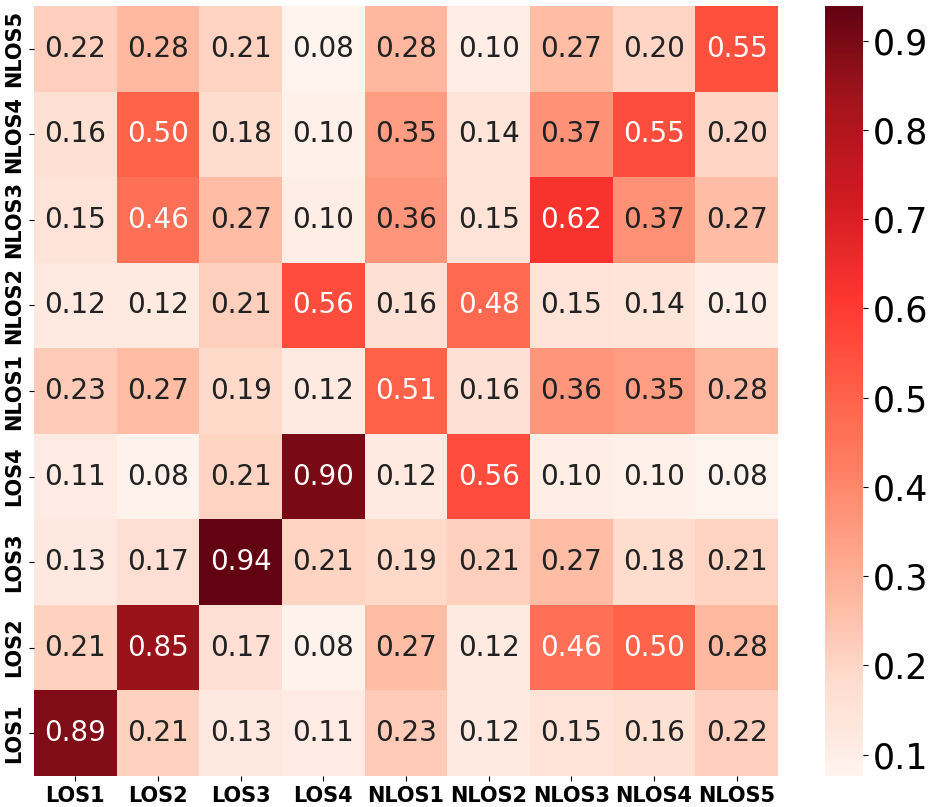}
            \caption{}
            \label{subfig:heatmap}
	\end{subfigure}
	\caption{(a) Topology of real-world dataset collection~\cite{dataset, web} and (b) Average inter-user channel correlation among clusters reproduced from ~\cite{dataset}.}
	\label{fig:latency}
\end{figure}

To evaluate our schemes under realistic massive MIMO deployments, we leverage a publicly available real-world channel dataset~\cite{dataset} from the RENEW wireless testbed~\cite{renew}. This dataset encompasses practical considerations such as multipath reflections, hardware impairments, and noise. The raw traces, collected on the 2.4 GHz Wi-Fi ISM band with a 20 MHz bandwidth and 52 OFDM subcarriers, contain received 802.11 Long Training Symbols (LTS). Channel matrices were subsequently extracted through channel estimation. The dataset represents a scenario with a BS equipped with 64 antennas serving 225 single-antenna UEs positioned at various locations.
Channel measurements were conducted for both Line-of-Sight (LOS) and Non-Line-of-Sight (NLOS) propagation conditions. The LOS component comprises four clusters, while the NLOS component comprises five clusters. Each cluster approximates a circular shape with a diameter of approximately 4 meters. The specific locations of the massive MIMO BS and user clusters are depicted in Fig.~\ref{subfig:topo}.  Within each cluster, user channels were measured at 25 uniformly distributed locations and Fig.~\ref{subfig:heatmap} shows the average inter-user correlation among clusters. It is obvious that users within the same cluster exhibit high correlation, whereas those in different clusters show low correlation, particularly in LoS clusters.


In our comparative evaluation, we consider all schedulers introduced in~\S\ref{sec:method}: the Greedy algorithm, Greedy Plus (GP), optimal solution through Gurobi, DRO, DRS, and RS\_ES. For the RB-orthogonal case, the Greedy algorithm serves as a baseline adopted by state-of-the-art (SOTA) work~\cite{radiosaber}, while the Gurobi optimizer provides the upper performance bound. For RB-sharing, since the exponential combinatorial space associated with UEs, slices, and RBs makes it infeasible to obtain an optimal solution in large-size networks, RS\_ES serves as the baseline method for comparison. 
All experiment configurations are listed in Table.~\ref{tab:ex_config}. We establish two static channel correlation scenarios: high-correlated (HC) and low-correlated. In the high-correlated scenario, users within a slice are deliberately chosen from the same LoS cluster, resulting in high inter-user correlation (reported to exceed 0.8 in~\cite{dataset}). Conversely, for the low-correlated (LC) scenario, users are positioned in distinct clusters within a slice, which ensures that inter-user correlation remains below 0.2 in the low-correlated case. To comprehensively assess performance, we also define two distinct sets of SLAs (loose and tight) corresponding to each scenario. For mobility channels, we consider both slow-mobility (SM) such as pedestrian speed and fast-mobility (FM) scenarios. Following the methodology outlined in \cite{dataset}, since channels of different spots in each cluster are measured continuously, it allows us to simulate a user moving randomly at low speed within a cluster. Similarly, channels across different clusters can be used to emulate high-speed mobility. 

\begin{table}[]
\scriptsize
    \caption{Overview of implemented experiment configurations. (LC is static low-correlated scenario, HC is static high-correlated scenario, SM is slow-mobility, FM is fast-mobility and PF is Proportional Fairness)}
    \begin{tabular}{c|c|c|l|c|c|c}
    \hline
    \textbf{Network Size}            & \textbf{Scenario} & \textbf{BS} & \textbf{RB} & \textbf{UE (Physical Clusters)}          & \textbf{Slice} & \textbf{K} \\ \hline
    \multirow{2}{*}{Small Size}      & LC                & 64          & 52          & 16 (4 LoS)           & 4              & 3, 8       \\ \cline{2-7} 
                                     & HC                & 64          & 52          & 16 (4 LoS)           & 4              & 3, 8       \\ \hline
    \multirow{3}{*}{Medium Size}     & LC                & 64          & 52          & 80 (4 LoS)           & 8              & 8, 16      \\ \cline{2-7} 
                                     & HC                & 64          & 52          & 80 (4 LoS)           & 8              & 8, 16      \\ \cline{2-7} 
                                     & HC + PF               & 64          & 52          & 80 (2 LoS + 2 NLoS)  & 8              & 16         \\ \hline
    \multirow{4}{*}{Real-World Size} & LC                & 64          & 52          & 200 (4 LoS + 4 NLoS) & 8              & 16         \\ \cline{2-7} 
                                     & HC                & 64          & 52          & 200 (4 LoS + 4 NLoS) & 8              & 16         \\ \cline{2-7} 
                                     & SM                & 64          & 52          & 200 (4 LoS + 4 NLoS) & 8              & 16         \\ \cline{2-7} 
                                     & FM                & 64          & 52          & 200 (4 LoS + 4 NLoS) & 8              & 16         \\ \hline
    \end{tabular}
    \label{tab:ex_config}
\end{table}
\begin{table}[]
\scriptsize
\begin{tabular}{c|c|c}
\hline
\textbf{Network Size}                                                                          & \textbf{SLA Stringency} & \textbf{Throughput SLA (Mbps)}                                  \\ \hline
\multirow{2}{*}{Small-size network}                                                            & Loose             & {[}51.9, 46.2, 50, 53.8{]}                  \\ \cline{2-3} 
                                                                                               & Tight             & {[}90.4, 84.6, 88.5, 92.3{]}                         \\ \hline
\multirow{2}{*}{\begin{tabular}[c]{@{}c@{}}Medium and \\ Real-world size network\end{tabular}} & Loose             & {[}16.7, 46.4, 42.3, 51.7, 19.2, 50.5, 48.1, 53.6{]} \\ \cline{2-3} 
                                                                                               & Tight             & {[}55.8, 84.2, 80.8, 91.2, 57.7, 88.3, 86.5, 92.4{]} \\ \hline
\end{tabular}
\end{table}

\subsection{Performance Evaluation}
\label{sec:exp_per}
We evaluate the performance of our proposed schedulers through various network sizes, differing stringencies of SLAs, and diverse correlation scenarios. In 5G, SLAs are negotiated by enterprise and service-provider based on radio resource availability, CAPEX and type of services+. We set loose and tight SLAs based on channel dataset following the instructions of~\cite{SLA}. For a fair comparison, all algorithms receive identical channels generated from real-world traces as input. We assess scheduler performance using the average number of allocated RBs across all TTIs while meeting SLAs. This metric indicates an algorithm's ability to satisfy SLAs with minimal RB utilization. Error bars are also incorporated into each plot to visualize the standard deviation in allocated RBs across TTIs. First, we employ throughput as the SLA to evaluate scheduler performance in a static scenario. Subsequently, we demonstrate our scheduler's superiority to support a variety of scheduling policies and SLAs, while achieving strong performance in mobility scenarios. Finally, we evaluate the running time of the proposed schedulers, highlighting the ability of DRO and DRS to deliver scheduling decisions within milliseconds. 

\subsubsection{Small-Size Network.}   
\begin{figure}[t]
	\centering
	\begin{subfigure}{0.24\linewidth}
		\centering
		\includegraphics[width=0.95\linewidth]{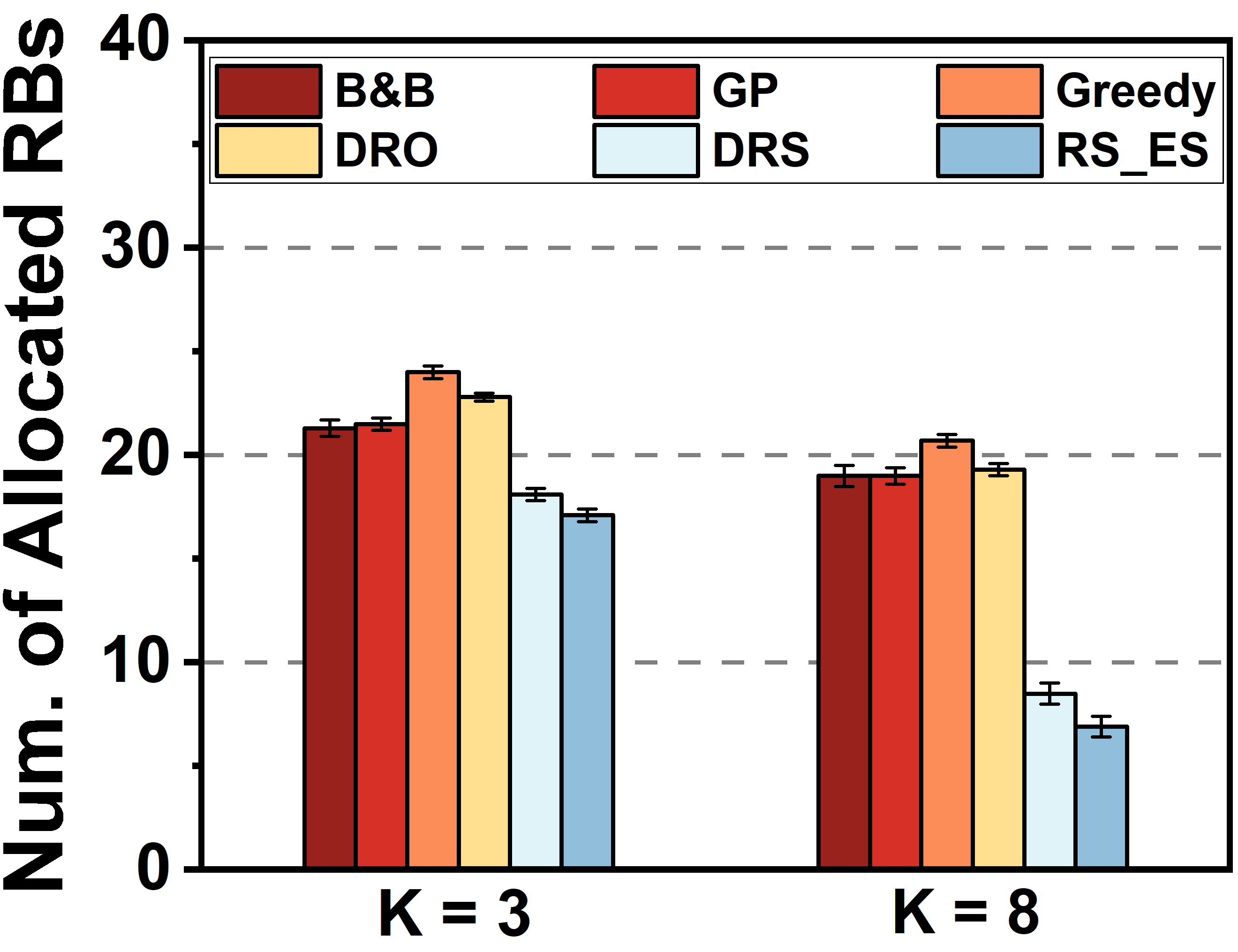}
            \caption{}
		\label{subfig:scl}
	\end{subfigure}
	\centering
	\begin{subfigure}{0.24\linewidth}
		\centering
		\includegraphics[width=0.95\linewidth]{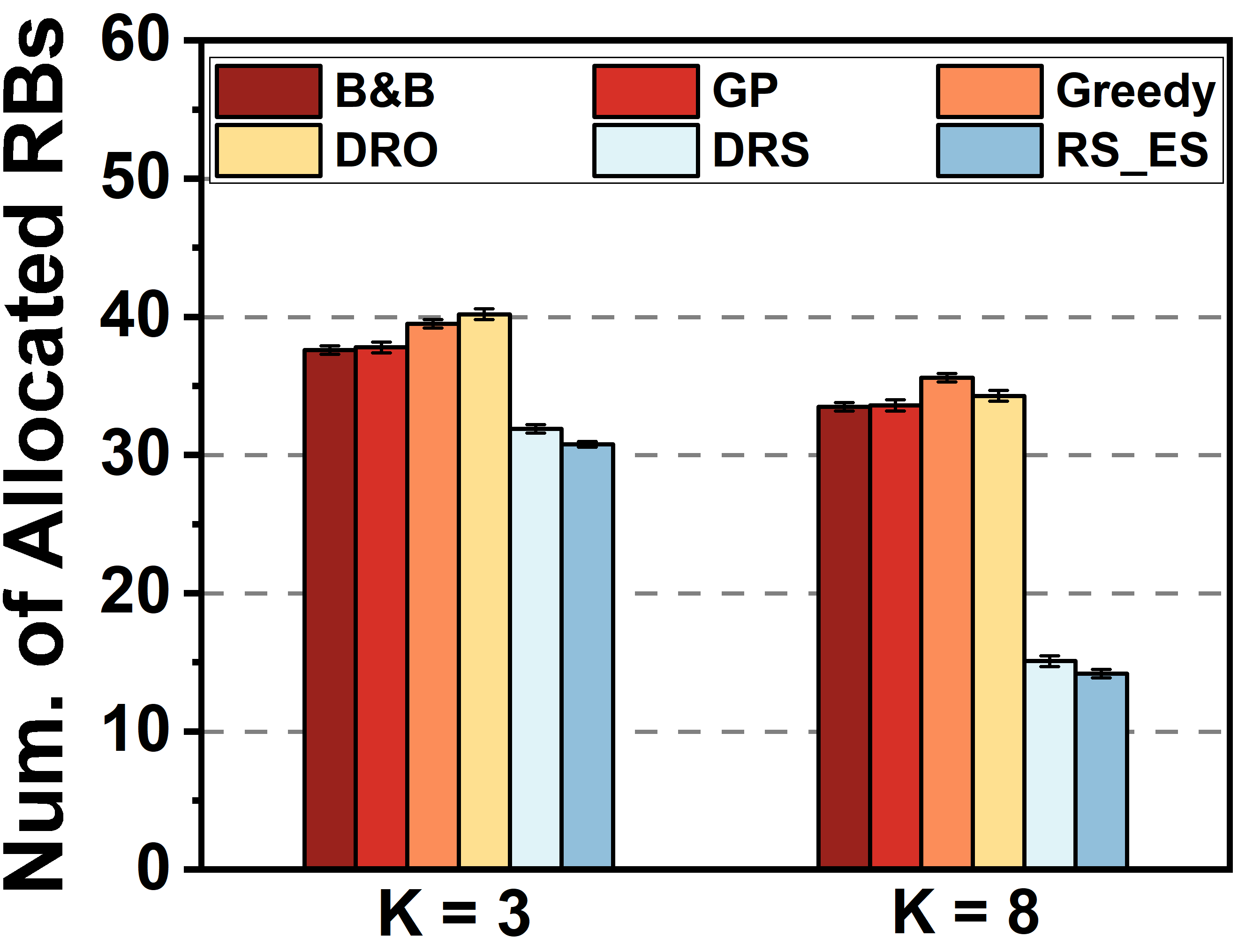}
            \caption{}
            \label{subfig:sct}
	\end{subfigure}
        \centering
	\begin{subfigure}{0.24\linewidth}
		\centering
		\includegraphics[width=0.95\linewidth]{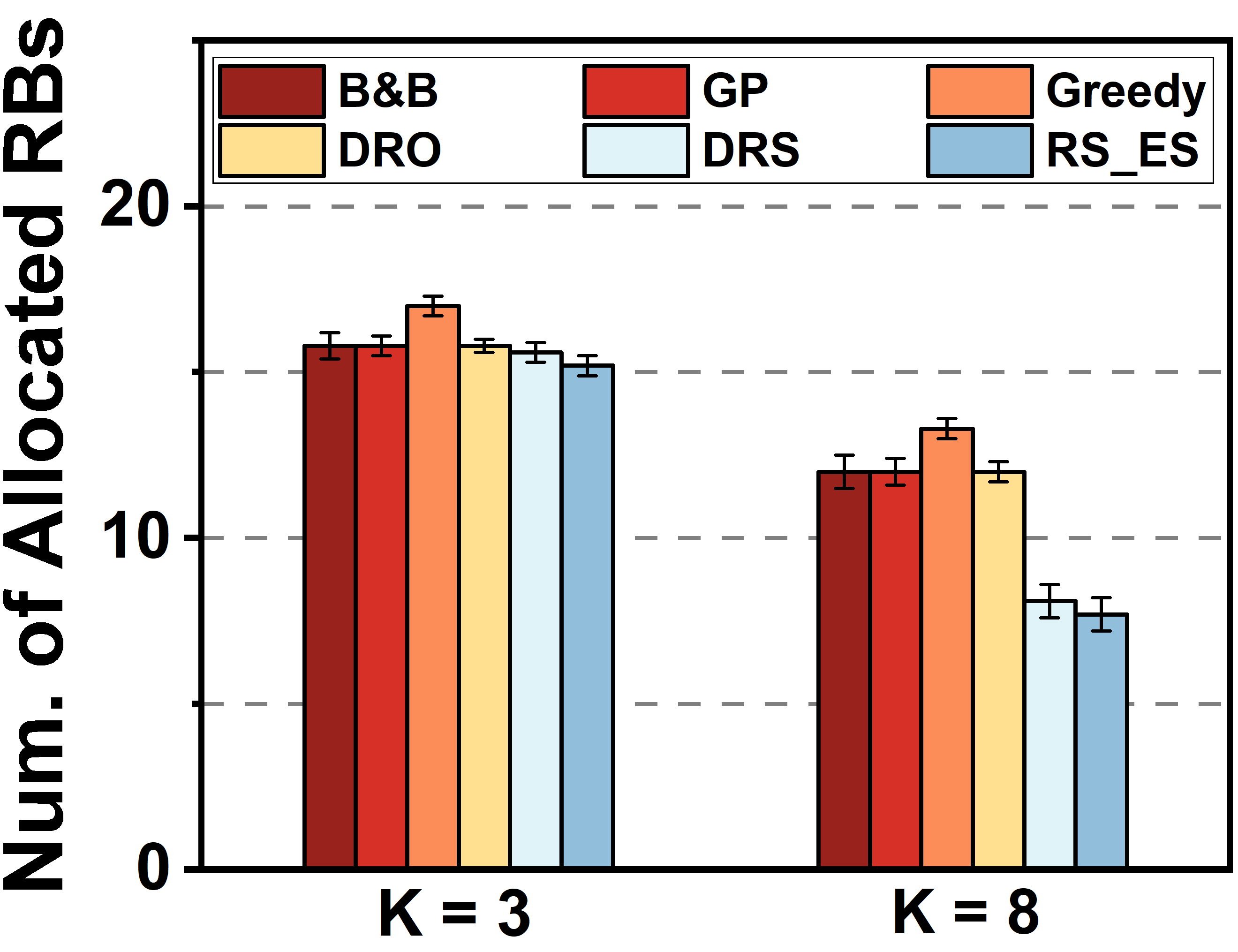}
            \caption{} 
            \label{subfig:sul}
	\end{subfigure}
	\centering
	\begin{subfigure}{0.24\linewidth}
		\centering
		\includegraphics[width=0.95\linewidth]{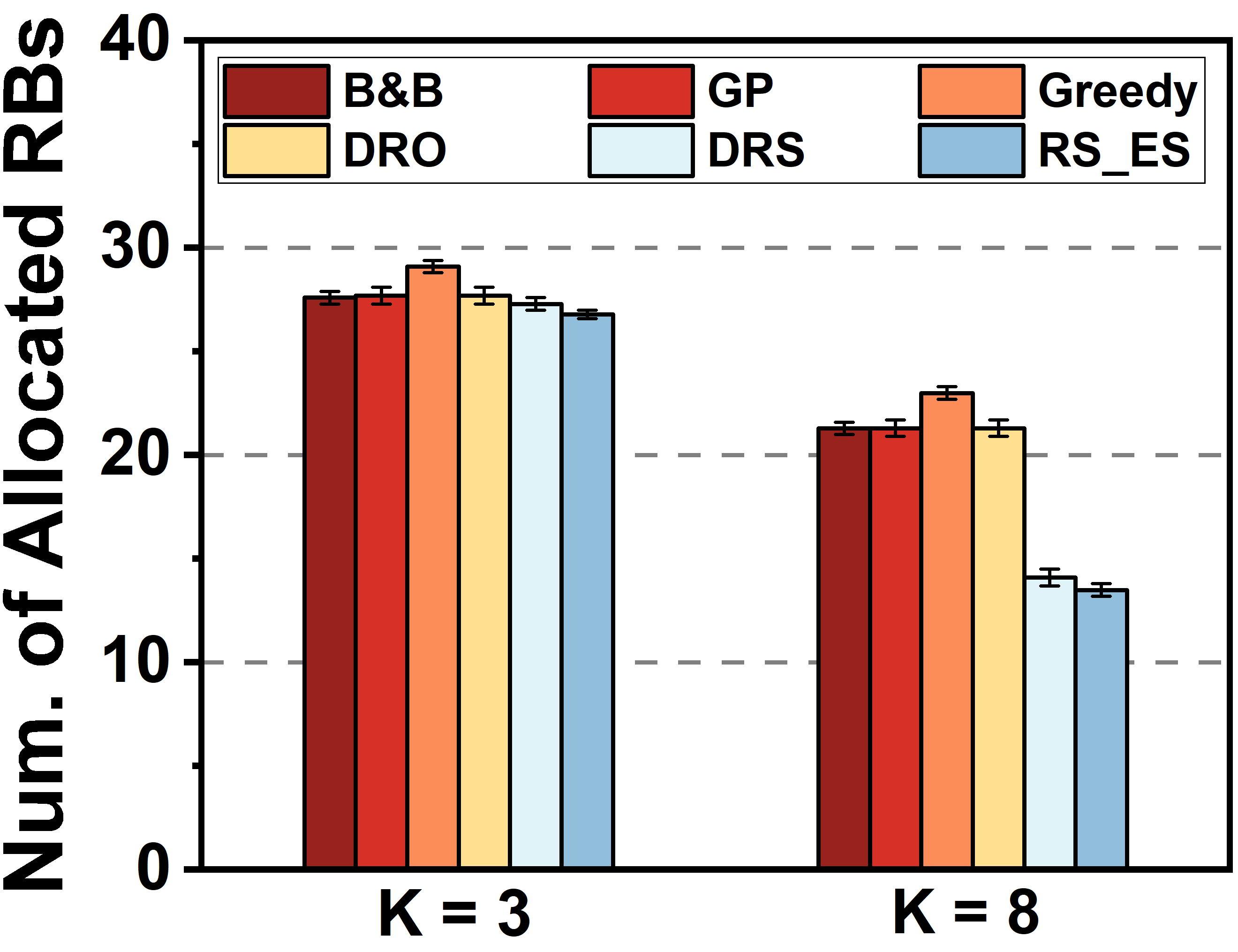}
            \caption{}
            \label{subfig:sut}
	\end{subfigure}
	\caption{Scheduler performance in small-size network (a) static high-correlated and loose SLA (b) static high-correlated and tight SLA (c)static low-correlated and loose SLA (d) static low-correlated and tight SLA.}
	\label{fig:small}
\end{figure}

We consider a massive MIMO network with 64 BS antennas, serving 16 users distributed across 4 slices, utilizing 52 RBs. Fig.~\ref{subfig:scl} and~\ref{subfig:sct} illustrate the number of RBs consumed under loose and tight SLA constraints within the high-correlated scenario. Evidently, GP outperforms Greedy in terms of RB utilization while being very close to the optimal solution obtained through B\&B. DRO exhibits slightly worse performance than GP (i.e. consuming one additional RB per TTI), but avoids the exhaustive search required by GP. Notably, DRS obtains near-optimal performance compared to RS\_ES and consumes much fewer RBs than any RB-orthogonal schedulers due the RB sharing opportunities among slices. Specifically, when $K=3$, DRS achieves the best performance with a reduction in RB consumption of 25\% and 19.2\% for loose and tight SLAs, respectively, compared to the Greedy algorithm. This advantage is further amplified to 58.9\% and 57.6\% when $K=8$. This behavior can be attributed to large $K$ emulating a user-sparse scenario ($\mathcal{K}_s < K$), where DRS can effectively share redundant resources with users from other slices, maximizing the utilization. Fig.~\ref{subfig:sul} and~\ref{subfig:sut} depict the performance of schedulers in the low-correlated scenario. In this case as well, DRS and DRO are able to achieve near-optimal performance in their respective scenarios. However, the advantage of RB sharing among slices is less pronounced compared to the high-correlated case. This is because, with low-correlated users, the impact of selecting users from the same slice (RB-orthogonality) or different slices (RB-sharing) is minimal. In contrast, for the high-correlated scenario, imposing RB-orthogonality introduces significant inter-user interference, leading to performance degradation. 

\subsubsection{Medium-Size Network.}
\begin{figure}[t]
	\centering
	\begin{subfigure}{0.24\linewidth}
		\centering
		\includegraphics[width=0.95\linewidth]{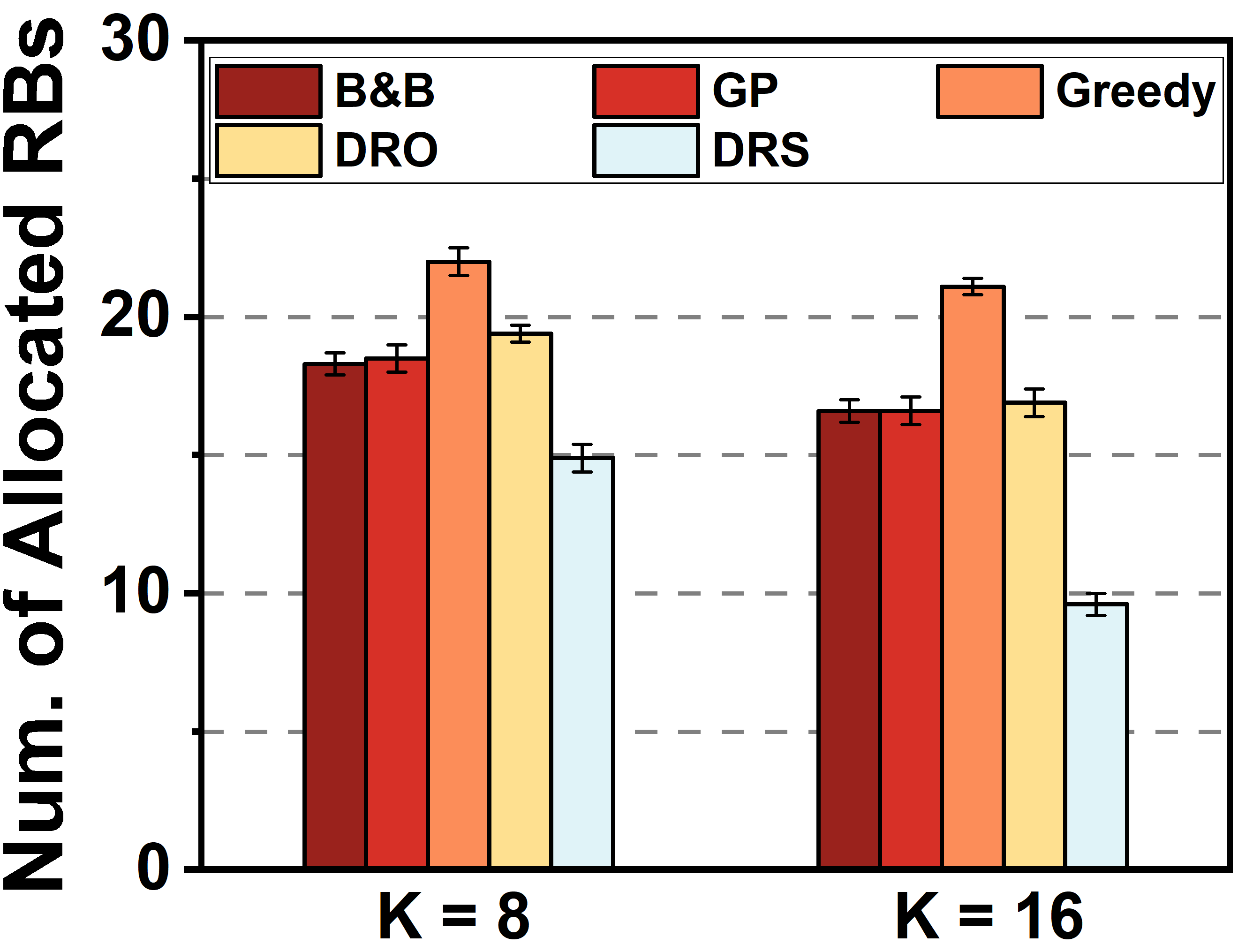}
            \caption{}
		\label{subfig:lcl}
	\end{subfigure}
	\centering
	\begin{subfigure}{0.24\linewidth}
		\centering
		\includegraphics[width=0.95\linewidth]{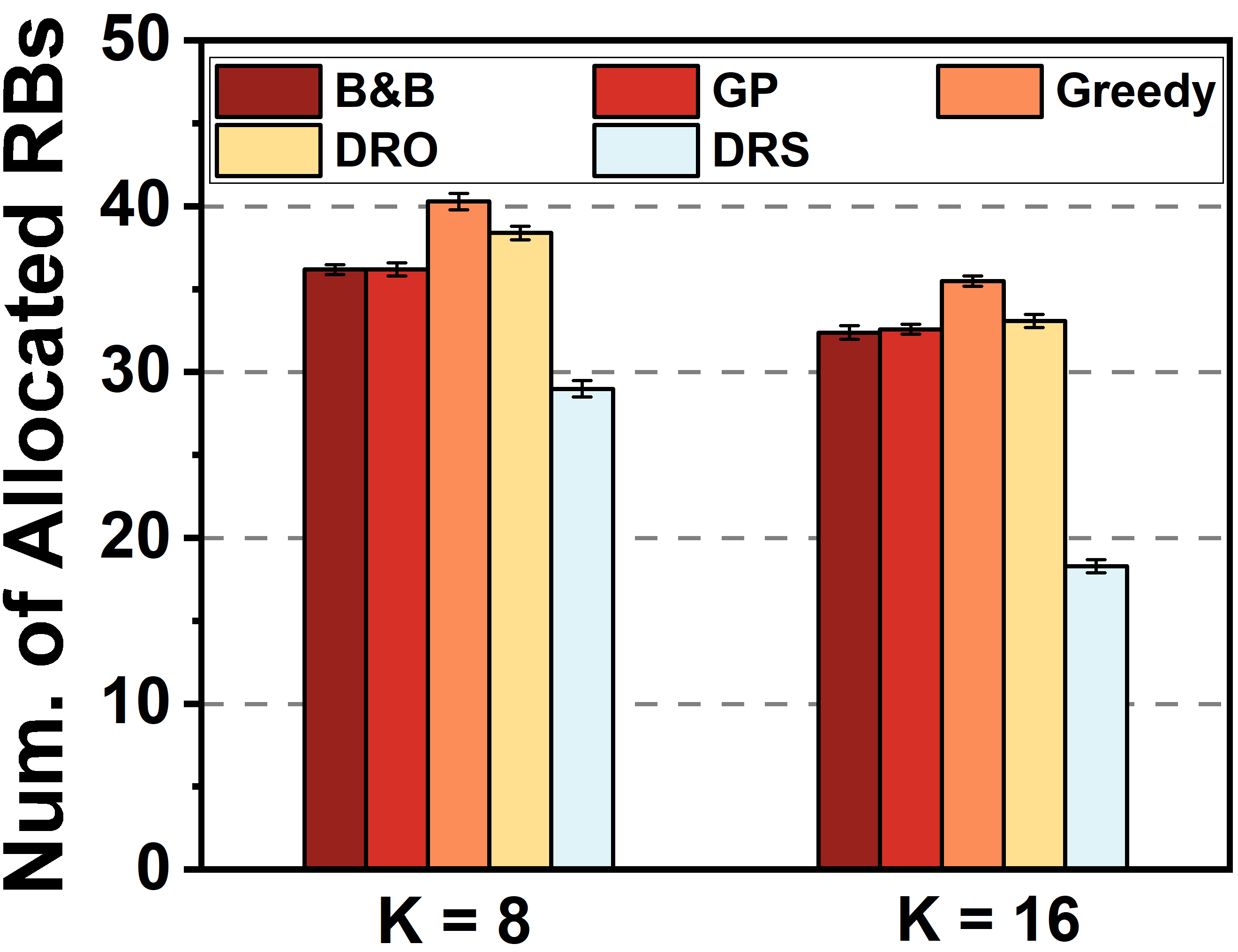}
            \caption{}
            \label{subfig:lct}
	\end{subfigure}
        \centering
	\begin{subfigure}{0.24\linewidth}
		\centering
		\includegraphics[width=0.95\linewidth]{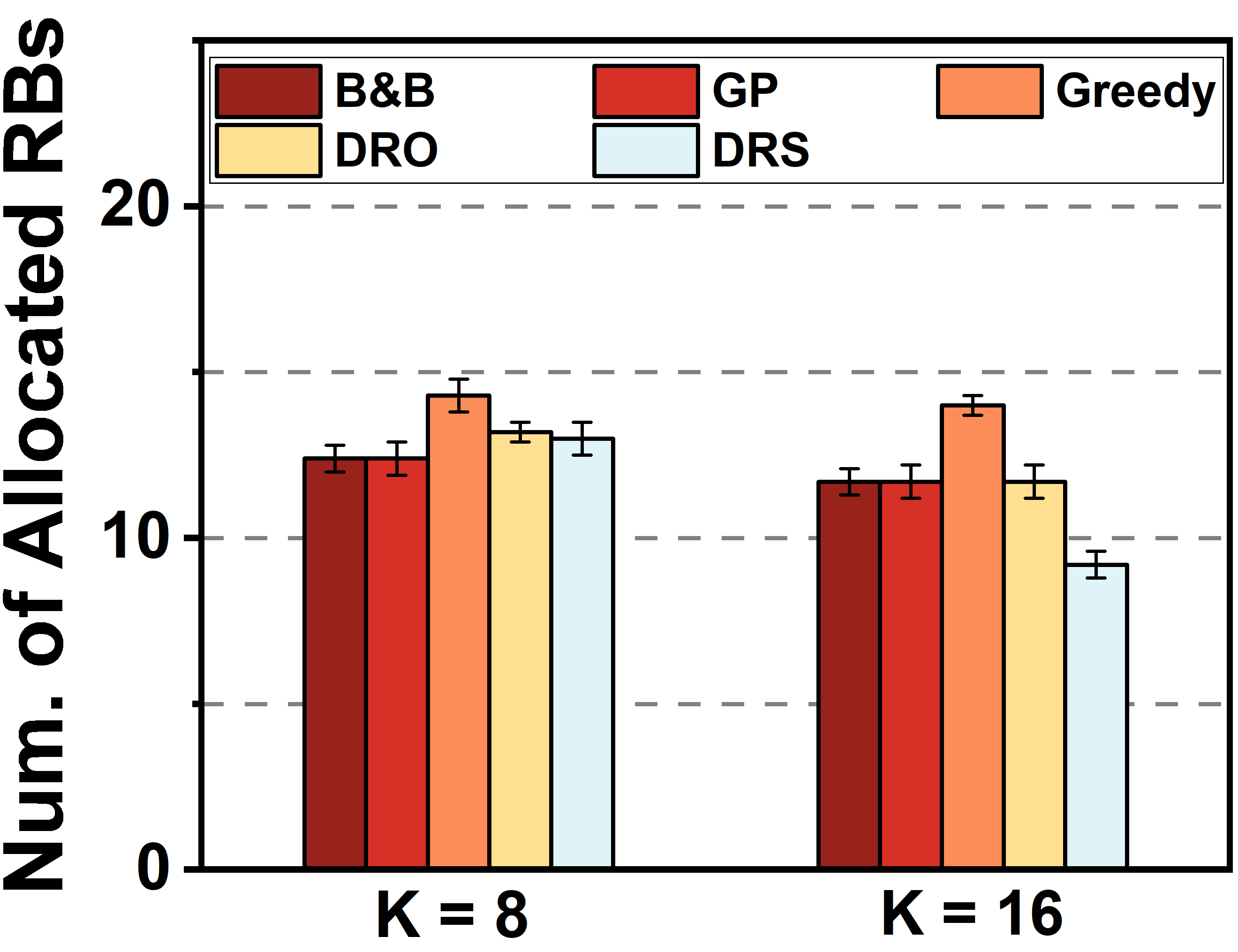}
            \caption{} 
            \label{subfig:lul}
	\end{subfigure}
	\centering
	\begin{subfigure}{0.24\linewidth}
		\centering
		\includegraphics[width=0.95\linewidth]{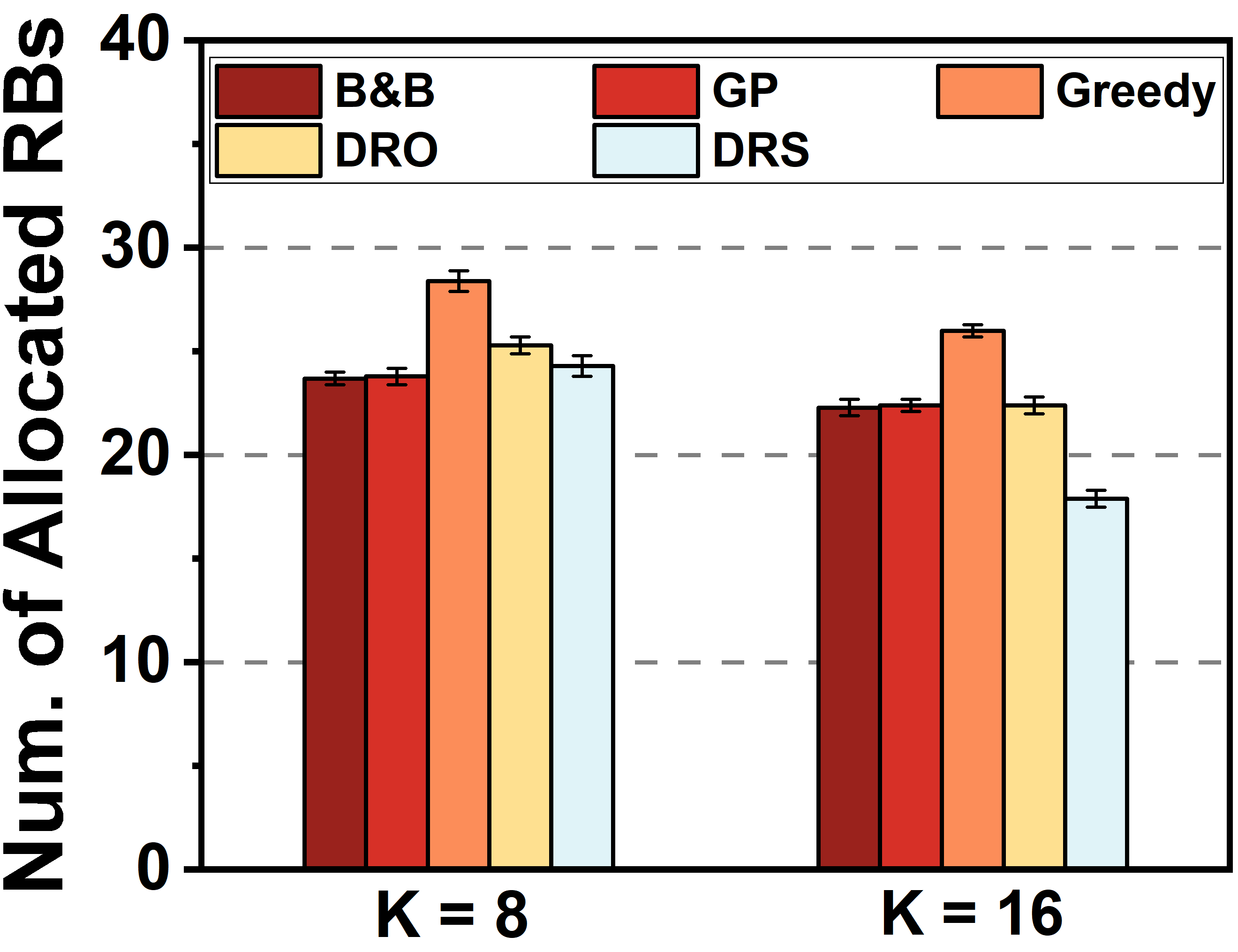}
            \caption{}
            \label{subfig:lut}
	\end{subfigure}
	\caption{Scheduler performance in medium-size network (a) static high-correlated and loose SLA (b) static high-correlated and tight SLA (c)static low-correlated and loose SLA (d) static low-correlated and tight SLA.}
	\label{fig:medium}
\end{figure}

We configure the network with 64 BS antennas, 52 RBs, and 80 users distributed into 8 slices, with each slice containing 10 users. 
However, RS\_ES is not implementable in this scale because of high complexity (i.e. $\mathcal{O}(\binom{80}{K})$). Fig.~\ref{fig:medium} reveals similar observations to those in the small-size network: (1) The performance of GP is close to optimal and surpasses the Greedy algorithm under any $K$, SLA constraints, and correlation conditions. (2) DRO performs slightly below GP due to its approximate method but with much less complexity. (3) DRS exhibits the advantage of RB sharing in a scalable and low complexity way, particularly with larger $K$ and in high-correlated scenario.

\subsubsection{Real-World-Size Network.} To conduct a realistic evaluation, we expand the network size to encompass 200 UEs with various channel conditions, originating from four LoS clusters and four NLoS clusters, while keeping all other configurations unchanged. Following the established topology of the clusters, we configure both high-correlated and low-correlated scenarios as previously applied to different network sizes. To make it more realistic, unlike the previous even distribution of users among slices, we generate a random set of numbers (i.e. [10, 12, 18, 20, 25, 33, 45, 37] with a mean of 25) to determine the number of users in each slice. With $K=16$, our experiments include both sparsely populated ($|\mathcal{K}_s| < K$) and densely populated ($|\mathcal{K}_s| > K$) slices. For this network size, the Greedy algorithm, GP, and Gurobi are infeasible due to their high complexity. Consequently, as depicted in Fig.~\ref{subfig:200r}, we only present the performance of DRO and DRS to show its scalability. 

\subsubsection{Mobility Scenarios.}
\label{sec:mob}
A comparison of the schedulers' performance in mobile scenarios is illustrated in Fig.~\ref{subfig:mob}. It is noteworthy that the results for the low-speed mobility scenario resemble those obtained in the static high-correlated scenario in Fig.~\ref{subfig:200r}, as users moving within the same cluster still remain highly correlated. Similarly, the results for high-speed mobility are comparable to those in the static low-correlated scenario, albeit with larger standard deviations than static scenario.

\subsubsection{Diverse Scheduling Policies.} 
\label{sec:policy}
\begin{figure}[t]
	\centering
	\begin{subfigure}{0.24\linewidth}
		\centering
		\includegraphics[width=0.95\linewidth]{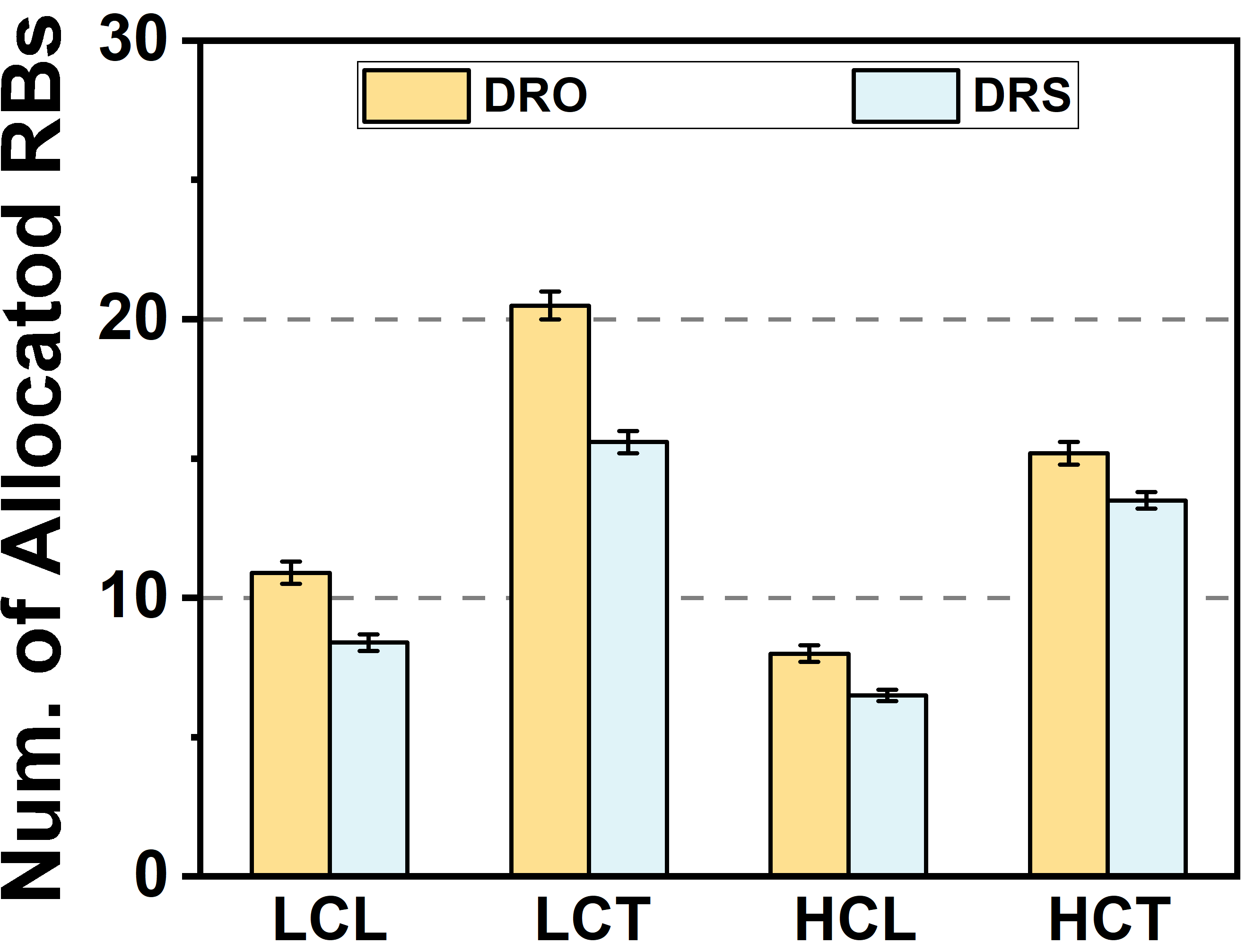}
            \caption{}
		\label{subfig:200r}
	\end{subfigure}
	\centering
	\begin{subfigure}{0.24\linewidth}
		\centering
		\includegraphics[width=0.95\linewidth]{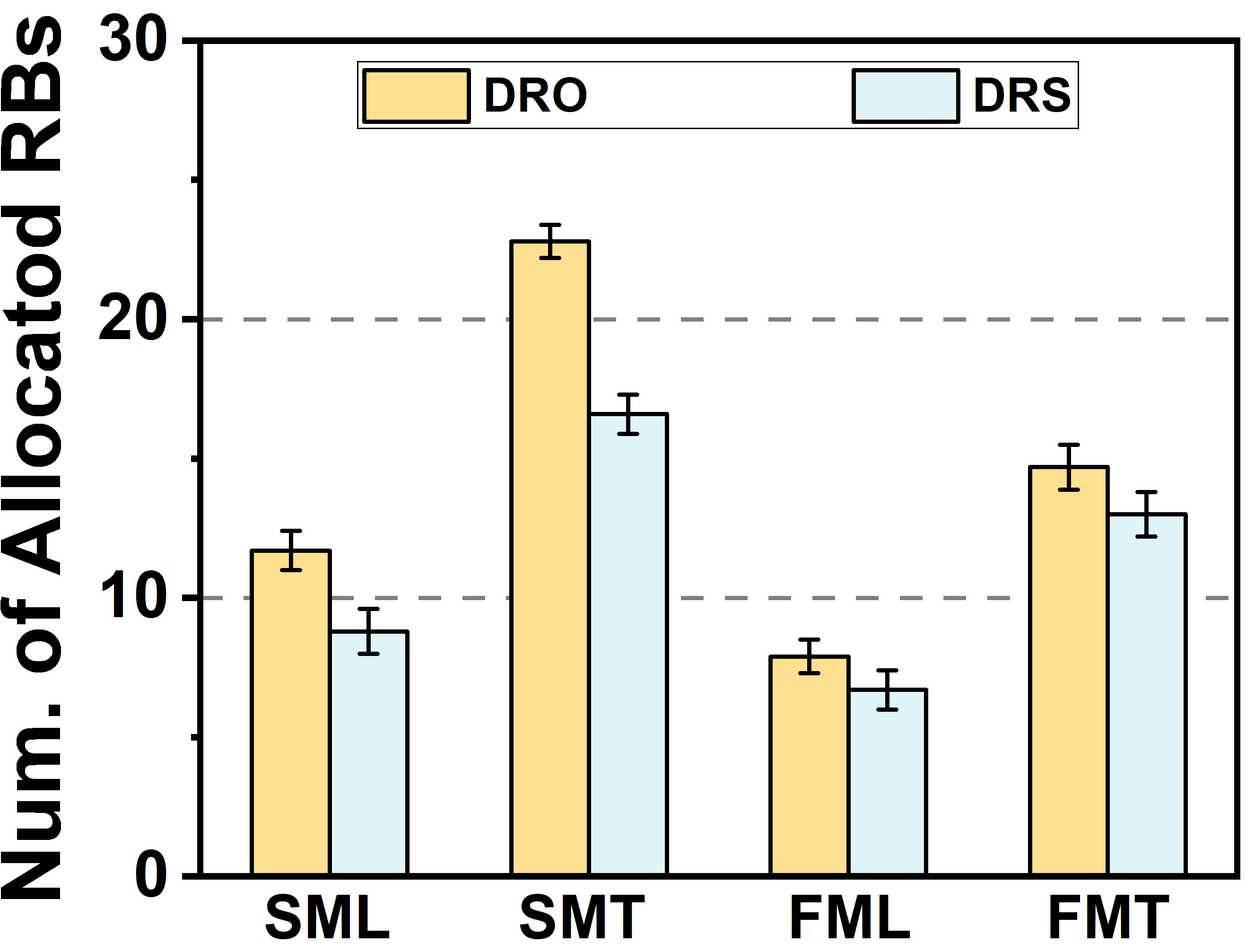}
            \caption{}
            \label{subfig:mob}
	\end{subfigure}
        \centering
	\begin{subfigure}{0.24\linewidth}
		\centering
		\includegraphics[width=0.95\linewidth]{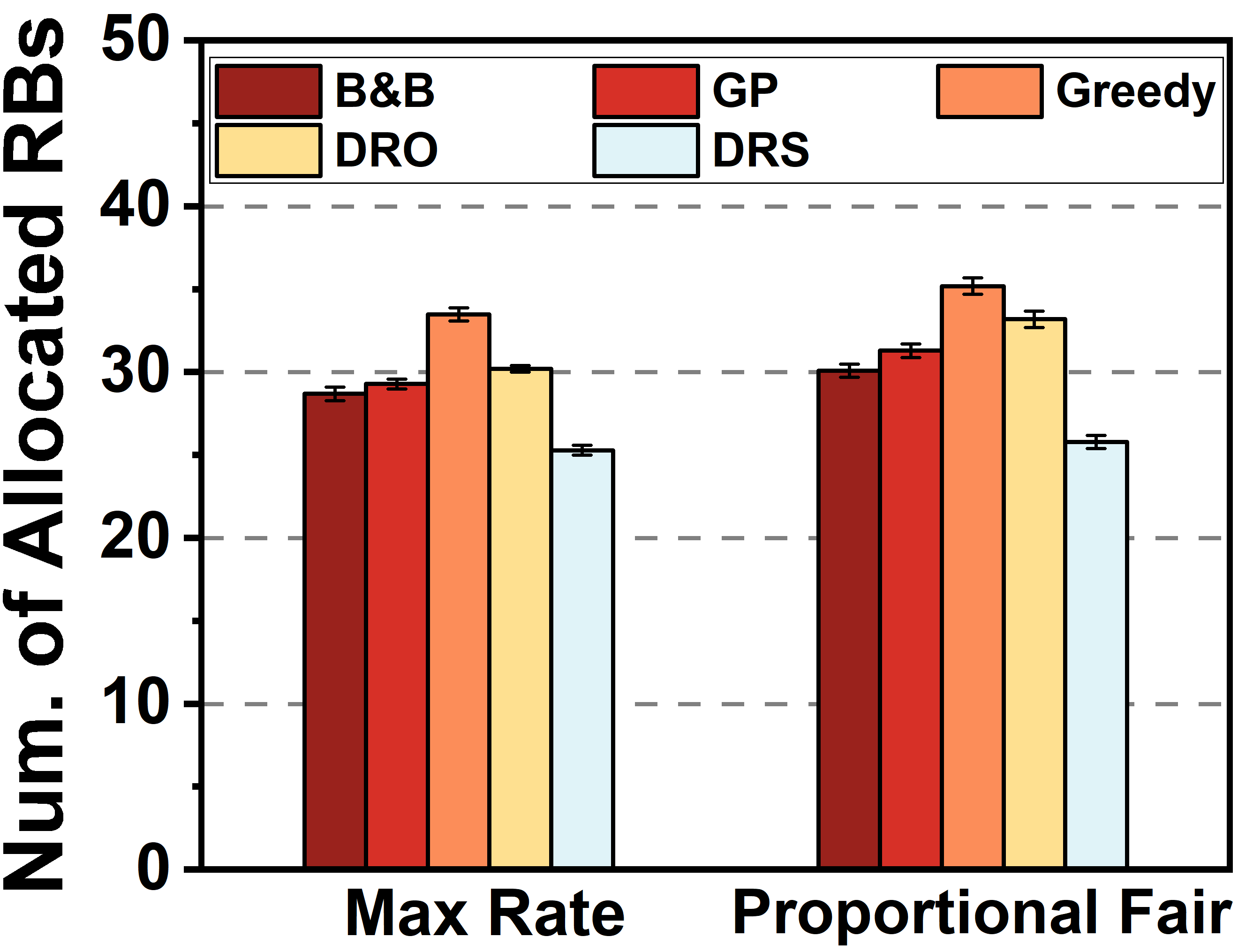}
            \caption{} 
            \label{subfig:pf_num}
	\end{subfigure}
	\centering
	\begin{subfigure}{0.245\linewidth}
		\centering
		\includegraphics[width=0.95\linewidth]{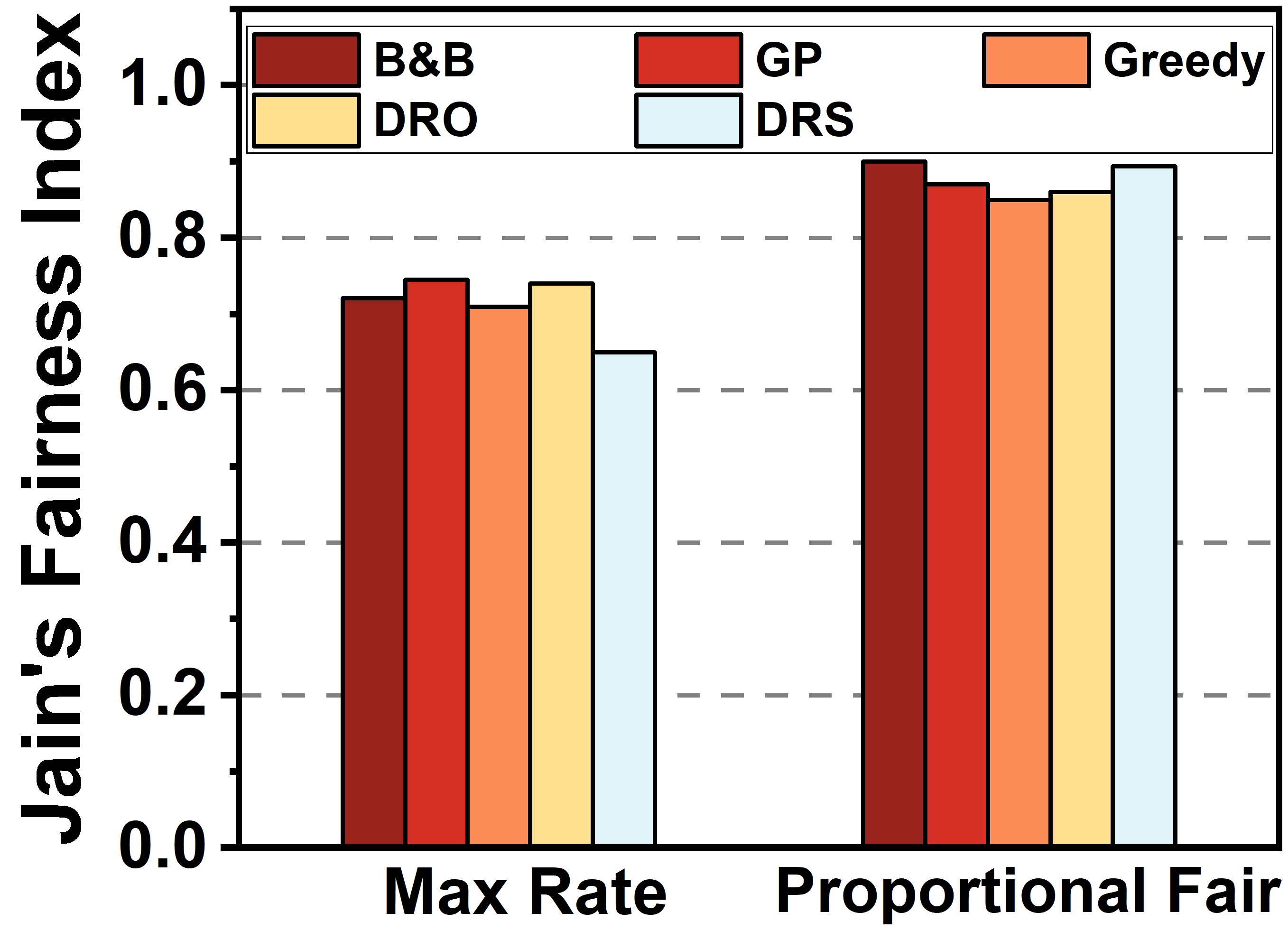}
            \caption{}
            \label{subfig:pf_jfi}
	\end{subfigure}
	\caption{Scheduler Performance: (a) and (b) real-world-size network with static and mobility scenario respectively. (c) Diverse scheduling polices in static medium-size network. (d) Jain's Fairness Index for Diverse scheduling polices in static medium-size network. (LCL:low-correlated with loose SLA, HCT: high-correlated with tight SLA, SML: slow-mobility with loose SLA, FMT: fast-mobility with tight SLA)}
	\label{fig:real}
\end{figure}

As illustrated in~\S\ref{sec:method}, channel gain serves as a key indicator of user channel quality and affects scheduling decisions to fulfill throughput SLAs. However, our schedulers can accommodate a variety of scheduling policies and SLA types by substituting channel gain with alternative performance metrics. In this evaluation, we utilize the proportional fairness metric in place of channel gain to implement the PF scheduling policy, thereby ensuring SLAs are met concerning both throughput and inter-user rate fairness. In this work, we employ Jain's Fairness Index (JFI)~\cite{jfi} to indicate rate fairness, which is a number between 0 to 1 and JFI=1 represents perfect fairness. PF metric is expressed as $\hat{g}^{b,t}_{k}/\hat{R}^{t}_{k}$, 
where $\hat{g}^{b,t}_{k}$ denotes normalized channel gain of user $k$ on RB $b$ at TTI $t$ and $\hat{R}^{t}_{k}$ indicates normalized accumulated rate of user $k$ by TTI $t$. It is essential to note that the normalization factors are the maximum channel gain and the maximum accumulated rate in $s$ where $slice(k) = s$. This is because only intra-slice fairness needs to be ensured, as guaranteeing global user fairness is impractical due to the varying SLAs across slices. To provide a thorough comparison, we conducted experiments in a medium-sized network with 80 UEs, incorporating the Greedy algorithm, GP, the optimal solution generated by the Gurobi optimizer (B\&B), DRO, and DRS. 
As illustrated in Fig.~\ref{subfig:pf_num} and~\ref{subfig:pf_jfi}, after replacing MR policy with PF, DRS is able to achieve an outstanding JFI but only uses additional 0.5 RBs in average per TTI compared to DRS with MR scheduling policy. Similarly, we can also incorporate other performance metrics to support various scheduling policies and corresponding SLAs. 

\subsubsection{Execution time for proposed algorithms}
\label{sec:exp_time}
As discussed in~\S\ref{sec:problem} and~\S\ref{sec:method},  both the Greedy algorithm and GP necessitate an exhaustive search by the intra-slice scheduler. This approach is feasible within SISO networks due to the availability of rate estimation via CQI utilizing a look-up table. However, in MIMO networks, 
computing rate for all user combinations within each slice on each RB is infeasible within a TTI. Conducting an exhaustive search on all $|B|$ RBs for a given slice $k$ containing $|\mathcal{K}_s|$ users incurs a time complexity of $\mathcal{O}(\binom{|\mathcal{K}_s|}{K}\times |B|)$. In contrast, approaches such as DRO or DRS compute rates only once after user selection on dedicated RBs, significantly reducing computational overhead. The primary computational bottleneck lies in the sorting of channel gains within each user group, detailed in~\S\ref{sec:method}, with a time complexity logarithmically dependent on the user group's size. Furthermore, as delineated in~\S\ref{sec:user_grouping}, user grouping can be executed in parallel on a massive scale, only requiring periodic execution per several TTIs.

\begin{figure}[t]
	\centering
	\begin{subfigure}{0.36\linewidth}
		\centering
		\includegraphics[width=0.95\linewidth]{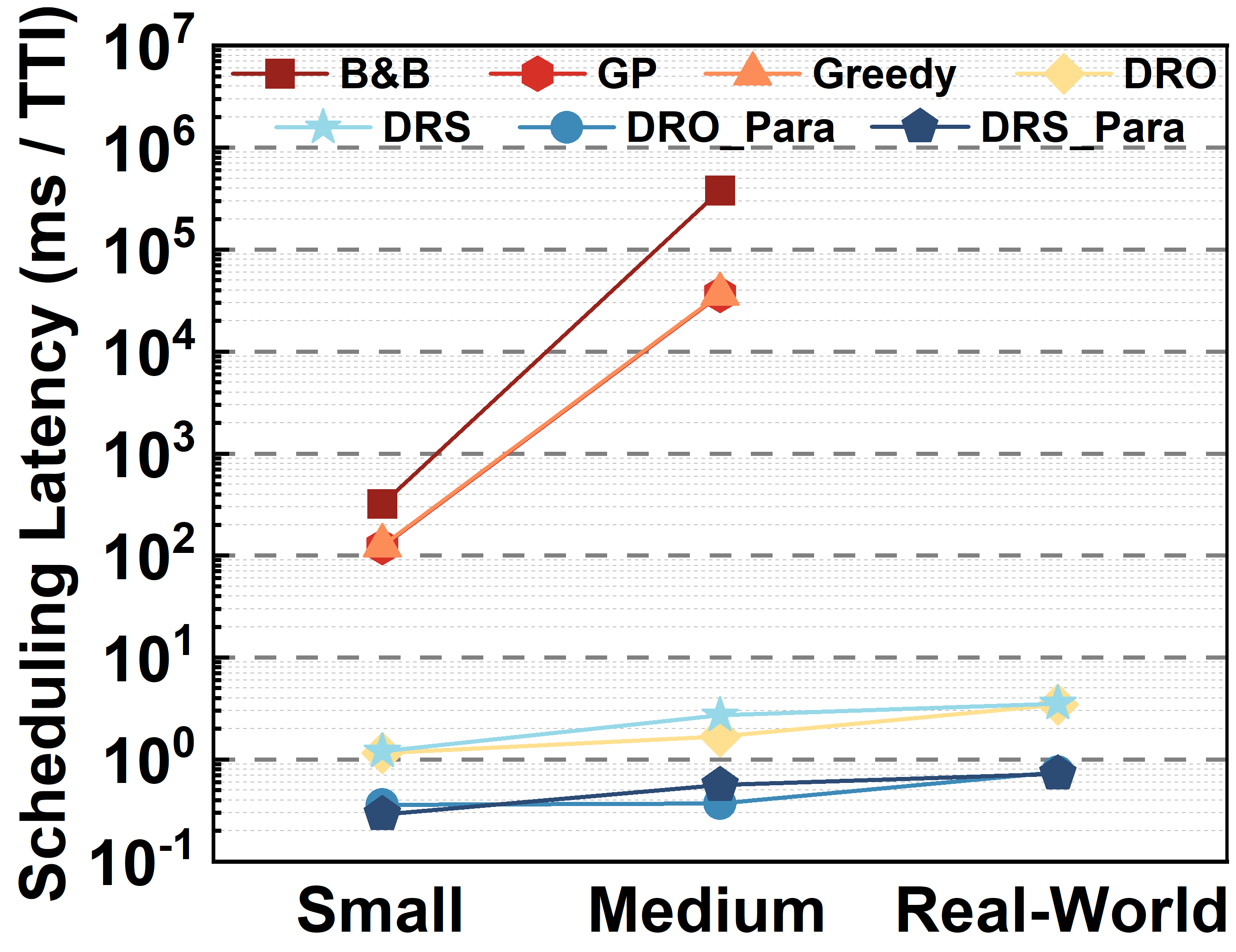}
            \caption{}
		\label{subfig:time}
	\end{subfigure}
	\centering
	\begin{subfigure}{0.36\linewidth}
		\centering
		\includegraphics[width=0.95\linewidth]{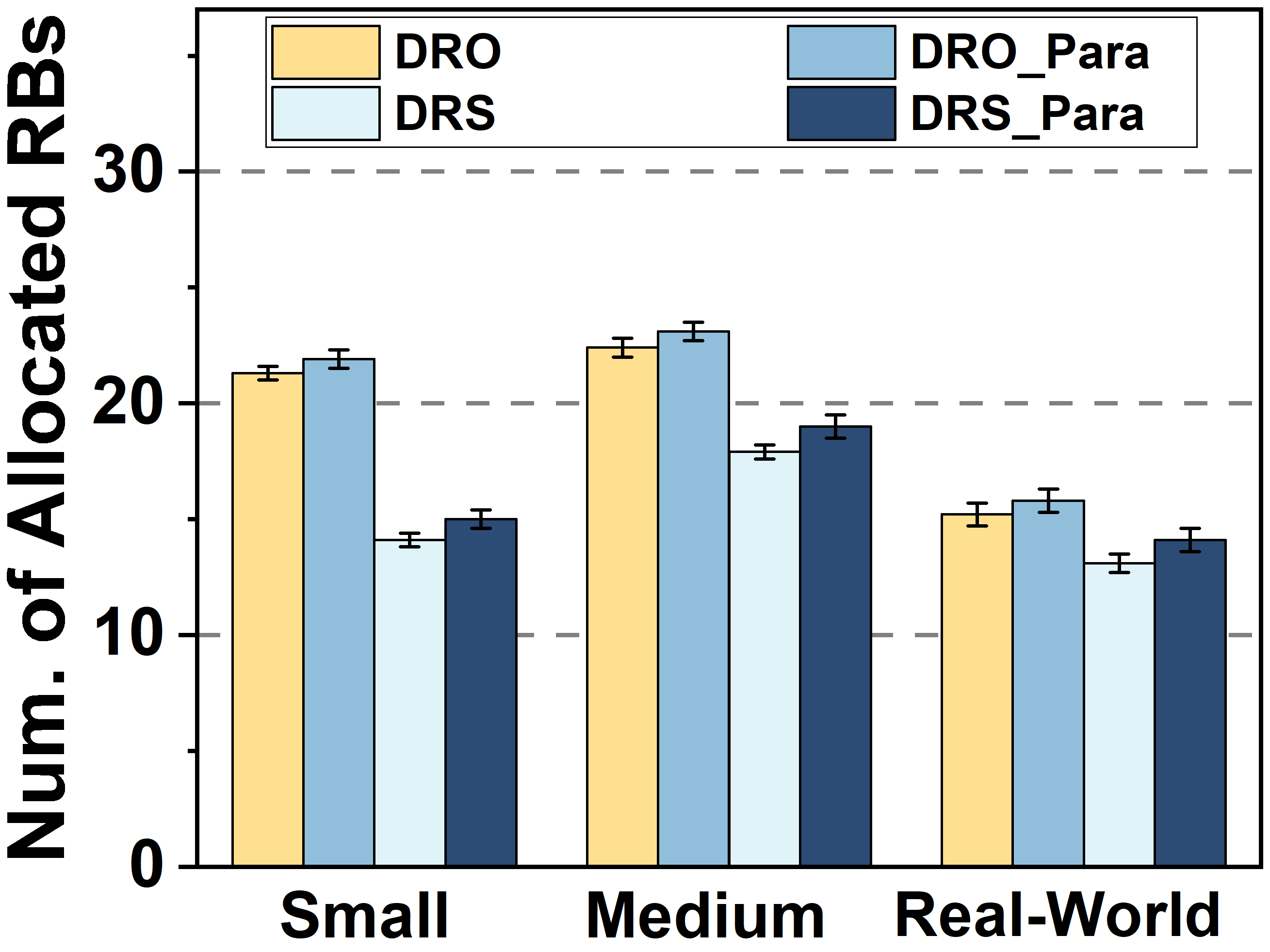}
            \caption{}
            \label{subfig:rb_latency}
	\end{subfigure}
	\caption{(a) Scheduling latency comparison in various sizes of networks and (b) RB consumption comparison between w/ and w/o RB parallelism.}
	\label{fig:latency}
\end{figure}

To evaluate the running time of proposed schedulers, we implement them using a single Intel Xeon core. Fig.~\ref{subfig:time} shows the scheduling time of GP and Greedy algorithm increases exponentially with the network size because of the exhaustive search that the intra-slice scheduler has to perform. B\&B does an exhaustive search at both inter-slice and intra-slice levels to find the optimal solution. Even though it has been accelerated by the Gurobi optimizer, it still takes the longest time to obtain scheduling decisions. In contrast, DRS and DRO grow linearly with the help of an approximate approach. However, as discussed in~\ref{sec:method}, because the outcomes of previous RB allocations adjust the SLA deficit, RB allocation in DRO and DRS occurs sequentially. It results in a scheduling time of slightly more than 1 ms for both DRS and DRO, which is unfavorable in real-world networks. Therefore, we adopt RB parallelism strategy as discussed in~\S\ref{sec:RB_parall} with DRS and DRO (labelled as DRS\_Para and DRO\_Para) to do allocation on multiple RBs in parallel. By doing this, DRS\_Para and DRO\_Para will spend a little more radio resources than DRS and DRO (i.e. up to 1.1 and 0.7 more RBs per TTI in average) in various network sizes but can significantly reduce the running time by 5 times so as to meet the stringent 5G latency requirement as Fig.~\ref{fig:latency} shown. 


\section{Limitations and Future Work}
\textbf{Timely and comprehensive channel information.} The proposed scheduler operates under the premise of obtaining precise and thorough channel data from users to do user grouping and estimate achieved rates for updating SLA-deficits. This assumption is prevalent in the majority of channel-aware RAN slicing literature~\cite{zipper, radiosaber, nvs, flare}. However, attaining such information proves challenging, particularly in large networks with numerous users. Notably, continuous updates to user grouping may be unnecessary, as it is a function of correlation statistics, which may exhibit minimal variation across different RBs and TTIs, as discussed in\S\ref{sec:user_grouping}. This insight suggests that our proposed scheduler has the potential to operate effectively with partial or slightly-outdated channel information, an aspect we aim to evaluate for its robustness in future work. 


\section{Conclusion}
\label{sec:conclusion}

This paper introduces a framework for RAN slicing aimed at resource scheduling in massive MIMO networks. The proposed scheduling scheme incorporates scheduling algorithms leveraging spatial-time slicing facilitated by beamforming in massive MIMO. In contrast to conventional multi-user MIMO schedulers focused on maximizing network throughput, our proposed scheduler addresses slice-level constraints for different enterprises catering to a group of distinct users, striving to minimize the number of RBs needed to meet each enterprise's SLA guarantees. The proposed schedulers encompass cooperative and exclusive resource allocation to enterprises. Depending on privacy requirements and traffic conditions, operators can opt to fulfill an enterprise's SLA through resource sharing with other enterprises via DRS or through exclusive resource allocation via DRO. With an extensive evaluation of all proposed schedulers using channels obtained from real-world massive MIMO test-bed measurements, the near-optimal performance of the proposed schemes across various scenarios, coupled with operational efficiency, are demonstrated.


\bibliographystyle{ACM-Reference-Format}
\bibliography{refs}


\end{document}